\documentclass{IEEEtran}

\usepackage{indentfirst}
\usepackage{url}
\usepackage{color}
\usepackage{footnote}
\usepackage[hang,flushmargin]{footmisc}
\usepackage[linesnumbered,boxed,ruled]{algorithm2e}
\usepackage{lipsum}
\usepackage{verbatim}
\usepackage[T1]{fontenc}
\usepackage[utf8]{inputenc}

\usepackage{amsmath}
\usepackage{amsfonts,amssymb}
\usepackage{amsthm}
\usepackage{bm}

\usepackage{graphicx}
\usepackage{epstopdf}
\graphicspath{{figures/}}
\usepackage{subfigure}
\usepackage{float}
\usepackage{rotating}
\usepackage{caption}
\usepackage{booktabs}
\usepackage{longtable}
\usepackage{multirow}
\usepackage{makecell}
\usepackage{array}
\usepackage{threeparttable}

\usepackage{mdwlist}
\usepackage{diagbox}
\usepackage{pifont}
\usepackage{pdflscape}

\newcommand{\Courier}{\fontfamily{pcr}\selectfont }
\renewcommand{\figurename}{Fig.~}
\renewcommand{\tablename}{Table~}
\definecolor{myRed}{RGB}{217,46,127}
\definecolor{myGreen}{RGB}{67,127,127}
\definecolor{MyResponse}{RGB}{0, 0, 0}

\title{LMD: A Learnable Mask Network to Detect Adversarial Examples for Speaker Verification}

\author{{Xing Chen, Jie Wang, Xiao-Lei Zhang, Wei-Qiang Zhang, and Kunde Yang}
\thanks{Corresponding author: Xiao-Lei Zhang}
\thanks{Xing Chen, Jie Wang, and Xiao-Lei Zhang are with the School of Marine Science and Technology, Northwestern Polytechnical University, Xi'an 710072, China, and with the Research \& Development Institute of Northwestern Polytechnical University in Shenzhen, China (e-mail: xing.chen@mail.nwpu.edu.cn, wangjie2017@mail.nwpu.edu.cn, xiaolei.zhang@nwpu.edu.cn). }
\thanks{Wei-Qiang Zhang is with the Department of Electronic Engineering, Tsinghua University, Beijing 100084, China (e-mail: wqzhang@tsinghua.edu.cn).}
\thanks{Kunde Yang is with the Ocean Institute of Northwestern Polytechnical University, China (e-mail: ykdzym@nwpu.edu.cn).}
}
\begin{document}
\maketitle

\begin{abstract}
Although the security of automatic speaker verification (ASV) is seriously threatened by recently emerged adversarial attacks, there have been some countermeasures to alleviate the threat. However, many defense approaches not only require the prior knowledge of the attackers but also possess weak interpretability. To address this issue, in this paper, we propose an \textit{attacker-independent} and \textit{interpretable} method, named \textit{learnable mask detector} (LMD), to separate adversarial examples from the genuine ones.
It utilizes score variation as an indicator to detect adversarial examples, where the score variation is the absolute discrepancy between the ASV scores of an original audio recording and its transformed audio synthesized from its masked complex spectrogram. A core component of the score variation detector is to generate the masked spectrogram by a neural network. The neural network needs only genuine examples for training, which makes it an attacker-independent approach. Its interpretability lies that the neural network is trained to minimize the score variation of the targeted ASV, and maximize the number of the masked spectrogram bins of the genuine training examples.
Its foundation is based on the observation that, masking out the vast majority of the spectrogram bins with little speaker information will inevitably introduce a large score variation to the adversarial example, and a small score variation to the genuine example.
Experimental results with 12 attackers and two representative ASV systems show that our proposed method outperforms five state-of-the-art baselines. The extensive experimental results can also be a benchmark for the detection-based ASV defenses.
\end{abstract}


\section{Introduction}
\IEEEPARstart{A}{utomatic} speaker verification (ASV) is a task of verifying the identity of a speaker by his (or her) pre-recorded utterance clips \cite{bai2021speaker}. Deep-learning-based speaker verification techniques can be categorized into two representative frameworks: \textit{stage-wise} \cite{variani2014deep,snyder2018x,liu2019large} and \textit{end-to-end} \cite{wan2018generalized,bai2020partial, chung2020defence}. A fundamental difference between the two frameworks is their loss functions, which are called \textit{classification-based} loss and \textit{verification-based} loss, respectively \cite{bai2021speaker}.
Both of the two frameworks have achieved excellent performance and have penetrated our daily lives with real-world applications such as authentication, bank transaction and forensics. However, adversarial attacks \cite{szegedy2013intriguing} were found to be able to defeat an ASV system even at a high signal-to-noise ratio (SNR) \cite{villalba2020x}, which brought great challenges to the applications of the ASV systems.

Adversarial attack is a technique that aims to induce an ASV system to make wrong decisions by adding human-imperceptible perturbations into the clean speech during the inference phase of ASV. The perturbed speech, a.k.a \textit{adversarial examples}, has been extensively studied in the ASV research \cite{lan2022adversarial,tan2022adversarial}. It can be generally classified into white-box attacks and black-box attacks. In the case of white-box attacks, i.e. the scenarios where the victim ASV model exposes all knowledge, including parameters, structure, and training data, to the attacker. Villalba \textit{et al.} \cite{villalba2020x} found that the state-of-the-art (SOTA) x-vector ASV models are extremely vulnerable even at a high SNR level of 30dB. Xie \textit{et al.} \cite{xie2021enabling} proposed to train a generator to efficiently craft adversarial examples. Since the white-box attacks have many obstacles in the reality, the black-box counterparts, which are knowledge-independent, have been paid more attention. Chen \textit{et al.} \cite{chen2021real} deployed a gray-box attack using only the output similarity scores of ASV. Further, ASV models were found to be vulnerable to transfer-based black-box attacks across training datasets \cite{kreuk2018fooling} and model structures \cite{li2020adversarial}. In addition, the works in \cite{xie2021enabling,wang2020inaudible} explored robust adversarial examples in terms of the universality and imperceptibility, respectively.
There are also some works focusing on applying adversarial attacks to realistic scenarios, such as the over-the-air \cite{li2020advpulse,zhang2021attack} or streaming input \cite{li2020advpulse,xie2021real} situations, and defeating the tandem system of ASV and its auxiliary subsystems \cite{zhang2021attack,zhang2022towards}.

Since adversarial attacks have posed the serious threat, it becomes foremost important to develop an effective countermeasure to protect the ASV systems. The current countermeasures fall into two categories: \textit{proactive defense} and \textit{passive defense}. Proactive defense mainly utilizes adversarial data augmentation techniques to retrain the ASV model, which is inconvenient to deploy \cite{lan2022adversarial}. For example, the works in \cite{wang2019adversarial,wu2020defense,pal2021adversarial} proposed to use adversarial examples generated by fast gradient sign method (FGSM), projected gradient descent (PGD) and feature scattering, respectively, to perform adversarial training defenses \cite{goodfellow2014explaining}. Passive defense does not modify the ASV model, instead, it defends against adversarial attacks by a mitigation or detection component. For example, the works in \cite{zhang2020adversarial,joshi2021study,wu2021improving} proposed to remove the adversarial noise with the adversarial separation network, Parallel-Wave-GAN (PWG) module, and cascaded self-supervised learning based reformer (SSLR), respectively. Wu \textit{et al.} \cite{wu2021voting} also employed a voting strategy with random sampling to mitigate the adversarial attacks.

This paper focuses on the detection-based passive defense approaches. There have been many works in this direction. Li \textit{et al.} \cite{li2020investigating} and Joshi \textit{et al.} \cite{joshi2022advest} discriminated adversarial and genuine examples by training a VGG-like binary classification network and an embedding feature extractor, respectively. However, their performance drops dramatically in unseen attacks, since the training relies on the prior knowledge of adversarial examples. Wu \textit{et al.} \cite{wu2021improving} picked out adversarial examples by the statistics of the similarity scores between enrollment utterance and synthesized utterances from multiple cascaded SSLRs. However, their experiments were conducted on the MFCC-level, which means it works in the time-frequency domain and relies on specific acoustic feature extractors. Peng \textit{et al.} \cite{peng2021pairing} proposed to train a binary classifier by the consistency of the scores of twin ASV models, i.e. a premier and a mirror one. Because training the classifier needs genuine examples only, the method gets rid of the dependence on specific attackers. However, it needs to find a SOTA fragile ASV and a rare robust ASV. Wu \textit{et al.} \cite{wu2022adversarial} proposed to detect adversarial examples by score variation, which was obtained by a vocoder composed of the Griffin-Lim (GL) algorithm or PWG model. However, it lacks strong interpretability, since there is no significant correlation between the training loss of PWG and the score variation in the detector. Chen \textit{et~al.} \cite{chen2022masking} separated adversarial examples from genuine ones by a masking operation at the feature-level. However, the masking operation is manually designed, and is dependent on the dimensionality of the input features.

To address the aforementioned issues of attacker-dependent, feature-dimensionality-dependent and manual selection, in this paper, we propose to detect adversarial examples by a \textit{learnable mask detector} (LMD). It takes score variation as an indicator, and calculates the score variation by a masking operation on complex spectrogram features. Specifically, it assumes that short-time fourier transform (STFT) disperses manually-added adversarial perturbation uniformly from the time domain to the time-frequency domain. Naturally, due to the robustness of the ASV model to noise, masking insignificant time-frequency bins of the complex spectrograms has a large impact on adversarial examples, and a small impact on genuine examples. Based on the above assumption and observation, we aim to learn an optimal mask matrix by a neural network, and then utilize the absolute discrepancy of ASV scores before and after the masking operation to detect the adversarial examples.

It is worth noting that (i) LMD only requires the genuine examples for training, so it is attacker-independent; (ii) LMD transforms the masked complex spectrograms to speech signals in the time domain by the inverse short-time fourier transform (iSTFT), thus it becomes feature dimensionality-independent; (iii) LMD obtains the mask matrix by a neural network automatically, instead of designed manually; (iv) further, LMD calculates the score variation of the detection as part of the training loss of the neural network, which makes the detection and training closely related.
We conducted experiments on two SOTA ASV models with diverse adversarial examples, and obtained an excellent detection performance. For example, detection equal error rates (EER) of no more than 5.9\% and 10.1\% are achieved on the ECAPA\_TDNN ASV and the Fast-ResNet34 ASV, respectively, in a noisy and blended detection scenarios.

Our contributions are summarized as follows:
\begin{itemize}
    \item We propose a mask-based and attacker-independent detector, named LMD, which effectively mitigated the threat posed by adversarial examples to ASV systems. To demonstrate the advantage of learning a mask matrix through a neural network as LMD, we also propose a manually designed masking complex spectrogram (MCS) method as a baseline.
    \item We conducted experiments on two SOTA ASVs with abundant attackers. The two ASVs, which behave as either victims or defenders, are derived from two representative frameworks, i.e. stage-wise ASV and end-to-end ASV. The attackers cover three kinds of generation algorithms, and act as either an impostor or an evader to the ASVs in both white-box and black-box attacks.
    \item Inspired by \cite{villalba2020x}, we evaluated the performance of a number of detectors under a given SNR budget. The experiments were also conducted in a scenario where the adversarial examples of a single attacker with different parameter settings were mixed, and the corresponding genuine examples were added with white-noise at the same SNR. Experimental results show that our proposed method outperforms the SOTA baselines in terms of the detection EER at an SNR budget of 37dB and the above.
\end{itemize}

The rest of the paper is organized as follows: Section~\ref{sec:preliminary} describes some preliminaries, including a brief introduction of ASV and three adversarial attack algorithms. Section~\ref{sec:methods} introduces our proposed methods. Section~\ref{sec:exp_setup} shows the experimental settings and evaluation metrics, while the results are discussed in Section~\ref{sec:exp_results}. Finally, Section~\ref{sec:conclusion} hands concluding remarks.

\section{Preliminaries} \label{sec:preliminary}
\subsection{Automatic Speaker Verification}
ASV aims to confirm whether an utterance is pronounced by a specified speaker. Deep-learning-based ASV consist of a speaker embedding extractor (including feature engineering, encoder network, and temporal pooling module), a training objective function, and a similarity scoring back-end \cite{bai2021speaker}. An encoder network first extracts frame-level speaker embeddings from acoustic feature sequences, e.g. logarithmic filter-banks (LogFBank). Then, segment-level speaker features are obtained by the cascading of a pooling module and a feed-forward network. Finally, either classification-based or verification-based objective functions are used to train the above frame-level and segment-level speaker embedding extractors jointly.

To demonstrate the generalizability of the proposed method to different ASV systems, we adopt two representative training objective functions, i.e. \textit{additive angular margin softmax} (AAM-Softmax) \cite{liu2019large} and \textit{angular prototypical} \cite{chung2020defence}, for the victim ASV systems. In the test phase of ASV, we determine whether a test utterance $\bm{x}^t$ and an enrollment utterance $\bm{x}^e$ belong to the same speaker by comparing the similarity of their speaker embeddings with a predefined threshold $\eta$. The test phase is formulated as:
\begin{equation}\label{equ:sv_calculate_score}
    s = \mathbf{S}\biggl(f\left(\bm{x}^t\right), f\left(\bm{x}^e\right);\, \bm{\theta}\biggr)\mathop{\gtrless} \limits_{H_0} ^{H_1} \eta ,
\end{equation}
where $\mathbf{S} \left(\cdot\, ;\, \bm{\theta}\right)$ denotes the well-trained ASV model $\mathbf{S}$ with parameters $\bm{\theta}$, $f\left( \cdot\right)$ is an acoustic feature extractor, and $H_1$ represents the hypothesis of $\bm{x}^t$ and $\bm{x}^e$ belonging to the same speaker, and $H_0$ is the opposite hypothesis of $H_1$, $s$ is the similarity score of the two embeddings. The higher the similarity score is, the more likely the hypothesis $H_1$ is true.

\subsection{Audio Adversarial Attack}
Audio adversarial attack refers to an emerging technique that artificially generates slight noise and blends it into genuine speech, so as to make a speech signal processing system behave wrongly according to the goal of the attacker \cite{szegedy2013intriguing}.

In terms of how much knowledge of the system is exposed to the attacker, we consider two attack scenarios: \textit{white-box} and \textit{black-box} attacks respectively. In the white-box attack scenario, the attacker has access to the full knowledge of the victim model, and can optimize the adversarial noise with the help of gradient from the victim model. In the black-box attack scenario, we consider the transfer-based cross-model attacker, who uses the adversarial examples generated by a substitute ASV model to attack the victim ASV model.

In terms of the goal of a attacker, we consider both \textit{impersonation} and \textit{evasion} types of attackers in this paper. There are two kinds of trials in a realistic ASV system, i.e. target trials and non-target trials. A target (or non-target) trial regards the test utterance $\bm{x}^t$ and the enrollment utterance $\bm{x}^e$ come from the same (or different) speakers. Therefore, there are two types of misclassification, which delivers two kinds of attackers: (i) a non-target trial is misclassified as a target trial, and (ii) a target trial is misclassified as a non-target trial. We refer to these two attackers as adversarial impersonation and adversarial evasion, respectively \cite{villalba2020x}. The adversarial impersonation (or evasion) aims to generate an adversarial test utterance, which will be judged by the victim ASV model as a target (or non-target) trial of the enrollment utterance.

In this paper, we employ two gradient-based attackers, which are the {basic iterative method} (BIM) \cite{kurakin2016adversarial} and PGD \cite{kurakin2016adversarial}, and an optimization-based attacker: Carlini Wanger (CW) \cite{carlini2017towards}, to craft adversarial example $\bm{\tilde{x}}^t$ for the test utterance $\bm{x}^t$. We describe each attacker in detail as follows.

\subsubsection{BIM}
It is an attacker that generates adversarial examples in a multi-step. At each iteration, it obtains the gradient of the similarity score with respect to the input utterance $\bm{x}_n$ and adds a perturbation of step $\alpha$ along the gradient direction, followed by a cropping operation. The BIM attacker searches an adversarial example via the following formula:
\begin{equation}\label{equ:bim_algorithm}
    \bm{x}_{n+1} = \operatorname{Clip}_{\bm{x}^t, \epsilon} \Bigl(\bm{x}_{n}+k \alpha \operatorname{sign}\bigl(\nabla_{\bm{x}_{n}}\mathbf{S}\left(\bm{x}_n;\, \bm{\theta},f\right)\bigr)\Bigr),
\end{equation}
where
\begin{equation*}
    k = \left\{\begin{array}{rl}
        1, & \text{if } \bm{x}^{e} \text{ and } \bm{x}^t  \text { contribute to a non-target trial} \\
        -1, & \text{if } \bm{x}^{e} \text{ and } \bm{x}^t \text { contribute to a target trial}
        \end{array}\right.
\end{equation*}
represents adversarial impersonation and adversarial evasion, respectively, and $n=0,1,\cdots,N$, with $N$ as the number of iterations, $\epsilon = N \alpha$ constrains the magnitude of the perturbation,  $\bm{x}_{n}$ is initialized by the test utterance, i.e. $\bm{x}_{0}=\bm{x}^t$ (note that, $\bm{x}^t$ is not normalized), $\operatorname{Clip}_{\bm{x}^t, \epsilon} \left(\cdot\right)$ denotes an element-wise clipping function which ensures the constraint $\|\bm{x}_{n} - \bm{x}^t\|_{\infty} \leq \epsilon$, and $\mathbf{S}\left(\cdot\,;\, \bm{\theta},f\right)$ denotes a function to calculate the similarity score in \eqref{equ:sv_calculate_score} when the enrollment utterance $\bm{x}^e$ is given. At the end of the $N$ iterations of the BIM attacker, an adversarial example $\bm{\tilde{x}}^t$ is found as $\bm{x}_{N}$.

\subsubsection{PGD}
It is essentially the same as BIM, but it initializes the perturbation to a random point in the $L_p$ norm ball and replaces the cropping operation in \eqref{equ:bim_algorithm} by the projection function. Instead of continuing to use the $L_{\infty}$ norm in BIM, we adopt its counterpart of $L_2$ norm in the PGD attacker to increase the diversity of the adversarial examples. The adversarial example $\bm{\tilde{x}}^t$ is also found as $\bm{x}_{N}$ via:
\begin{equation}\label{equ:pgd_algorithm}
    \bm{x}_{n+1} = \Pi_{\bm{x}^t + \mathcal{S}, \epsilon} \Bigl(\bm{x}_{n}+k \alpha \frac{\nabla_{\bm{x}_{n}}\mathbf{S}\left(\bm{x}_n;\, \bm{\theta},f\right)}{\|\nabla_{\bm{x}_{n}}\mathbf{S}\left(\bm{x}_n;\, \bm{\theta},f\right)\|_{2}}\Bigr),
\end{equation}
where $k$, $n$, $N$, $\alpha$ and $\epsilon$ are defined in \eqref{equ:bim_algorithm}, $\Pi_{\bm{x}^t + \mathcal{S}, \epsilon}\left(\cdot\right)$ represents a function of mapping the input into the sphere of $L_2$ norm, which ensures the constraint $\|\bm{x}_n - \bm{x}^t\|_{2} \leq \epsilon$.

\begin{table}[t]
    \setlength\tabcolsep{5pt}
    \setlength\extrarowheight{4pt}
    \newlength\lencell \newlength\lenbase \newlength\lengoal
    \settowidth{\lenbase}{six settings for $N$}
    \setlength{\lencell}{\dimexpr0.5\lenbase-\tabcolsep-\arrayrulewidth/2\relax}
    \setlength{\lengoal}{\dimexpr1.5\lenbase+\tabcolsep+\arrayrulewidth/2\relax}

    \newcolumntype{P}[1]{>{\centering\arraybackslash}p{#1}} 
    \newcommand\twocols[1]{\multicolumn{2}{P{\lenbase}}{#1}}  
    \newcommand\twocolsL[1]{\multicolumn{2}{|P{\lenbase}}{#1}}
    \newcommand\threecols[1]{\multicolumn{3}{P{\lengoal}}{#1}}
    \newcommand\threecolsL[1]{\multicolumn{3}{|P{\lengoal}}{#1}}

    \centering
    \caption{Twelve types of attackers adopted in this paper. Each of the attacker is composed by an algorithm, a type of prior knowledge, and an attack goal from the options listed in the table.}
    \scalebox{0.86}{\begin{tabular}{l|*{6}{p{\lencell}}}
    \toprule
    \multirow{2}{*}{Algorithm formulation} & \twocols{BIM ($L_{\infty}$)}    & \twocolsL{PGD ($L_2$)}            & \twocolsL{CW (RMS)} \\
     & \twocols{6 settings for $N$}  & \twocolsL{6 settings for $N$}   & \twocolsL{6 settings for $\kappa$} \\
    \midrule
    Prior knowledge  & \threecols{\textit{White-box} attack}    & \threecolsL{\textit{Black-box} attack}  \\
    \midrule
    Attack goals      & \threecols{\textit{Impersonation}}       & \threecolsL{\textit{Evasion}} \\
    \bottomrule
    \end{tabular}}
    \label{tab:attacker_types}
\end{table}

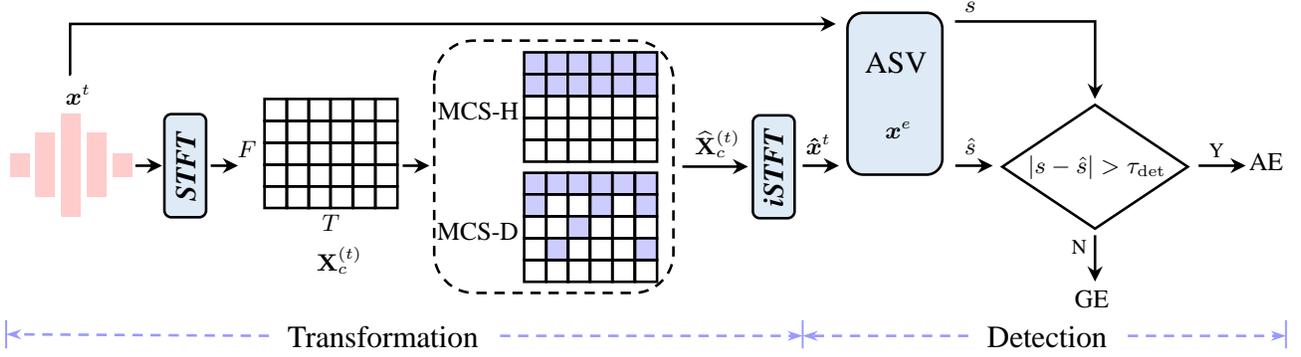
\begin{figure*}[t]
    \centering
    \includegraphics[width=0.95\textwidth]{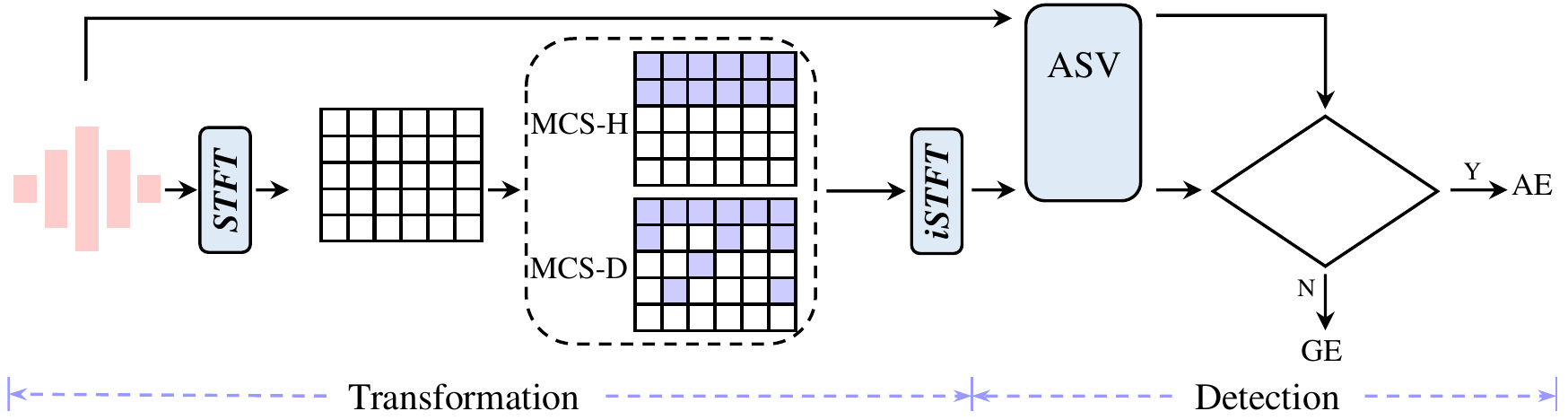}
    \put(-465,95){\small\bfseries\color{black}{$\bm{x}^t$}}
    \put(-400,75){\small\bfseries\color{black}{$F$}}
    \put(-368,47){\small\bfseries\color{black}{$T$}}
    \put(-370,32){\small\bfseries\color{black}{$\mathbf{X}_{c}^{\left(t\right)}$}}
    \put(-227,76){\small\bfseries\color{black}{$\mathbf{\widehat{X}}_{c}^{\left(t\right)}$}}
    \put(-185,76){\small\bfseries\color{black}{$\bm{\hat{x}}^t$}}
    \put(-155,82){\small\bfseries\color{black}{$\bm{x}^e$}}
    \put(-125,76){\small\bfseries\color{black}{$\hat{s}$}}
    \put(-125,130){\small\bfseries\color{black}{$s$}}
    \put(-101,69){\small\bfseries\color{black}{$|s-\hat{s}| > \tau_{\rm det}$}}

    \caption{Pipeline of the Masking Complex Spectrogram (MCS) detection method. The symbols $\bm{x}^t$ and $\mathbf{X}_{c}^{\left(t\right)}$ denote the original test utterance and its complex spectrogram features respectively, and $\bm{\hat{x}}^t$, $\mathbf{\widehat{X}}_{c}^{\left(t\right)}$ are the corresponding transformed ones. The ASV score variation $|s-\hat{s}|$ after the masking operation is used to identify whether the input utterance $\bm{x}^t$ is an adversarial example (AE) or a genuine example (GE).}
    \label{fig:mcs_detect}
\end{figure*}

\subsubsection{CW}
It is an optimization-based approach. It aims to get the minimum perturbation $\bm{\delta}^{*}$ for a successful attack and crafts an adversarial example by $\bm{\tilde{x}}^t=\bm{x}^t + \bm{\delta}^{*}$,
\begin{equation}\label{equ:cw_algorithm}
    \bm{\delta}^{*}=\operatornamewithlimits{min}_{\bm{\delta}}\frac{\|\bm{\delta}\|_{2}}{\sqrt{L}} + c \mathcal{J}\left(\bm{x}^t + \bm{\delta}\right),
\end{equation}
where $L$ is the length of the input test utterance $\bm{x}^t$, the normalized $L_2$ distance, a.k.a the root mean square (RMS) distance, of the perturbation is adopted to eliminate the effect of signal duration \cite{villalba2020x}, and $c$ is a hyperparameter to balance the imperceptibility and aggressiveness of the adversarial perturbation, which is found by a binary search procedure. The optimization objective of the aggressiveness $\mathcal{J}\left(\cdot\right)$ is defined as:
\begin{equation}\label{equ:cw_aggre_objective}
    \mathcal{J}\left(\cdot\right) = \left\{\begin{array}{lc}
        \! \max \Bigl(0, - \mathbf{S}\left(\cdot\,;\, \bm{\theta},f\right) + \left(\eta + \kappa\right) \Bigr), \!\! & \text{impersonation }  \\
        \vspace{-10pt} \\
        \! \max \Bigl(0, \mathbf{S}\left(\cdot\,;\, \bm{\theta},f\right) - \left(\eta - \kappa\right) \Bigr),\!\! & \text{evasion }
        \end{array}\right.
\end{equation}
where $\eta$ is a decision threshold and $\kappa$ is a confidence value.

Finally, we summarize the adversarial attackers to the two ASV models that will be used in this paper as in \tablename{\ref{tab:attacker_types}}, which covers most types of attacks in literature.

\section{Methods} \label{sec:methods}

In this section, we first present the motivation of the proposed method in Section~\ref{ssec:motivation}, then present the framework of the proposed method in Section~\ref{subsec:framework}, and finally present two implementations of the framework in Sections \ref{subsec:mcs} and \ref{subsec:lmd} respectively.

\subsection{Motivations} \label{ssec:motivation}
Although adversarial examples seriously threaten the security of ASV, detection-based adversarial defense methods can effectively alleviate this threat. Based on the assumption that adversarial perturbations are uniformly distributed in acoustic features, Chen \textit{et al.} \cite{chen2022masking} proposed Masking LogFBank features (MLFB) to detect adversarial examples. More specifically, masking as many insignificant speech features as possible will have a small impact on genuine examples and a large impact on adversarial examples, and thus utilize the variation of similarity scores after the masking operation to detect adversarial examples. However, MLFB has two problems: (i) \textit{Non-universal}. Since MLFB performs masking operation directly on the input feature of an ASV system, its manually selected threshold is related to the dimensionality of the input feature. Moreover, when the dimensionality of the input feature decreases, which means the granularity of the features becomes coarser, MLFB may fail. (ii) \textit{Hand-crafted mask}. MLFB masks the time-frequency bins of the input feature, either at high frequencies (MLFB-H) or using one-order difference (MLFB-D), both of which rely on human experience and lead to sub-optimal detection performance.

To address the above two shortcomings, we make improvements from two aspects respectively. For the non-universal problem, we perform ideal binary masking (IBM) operation on the complex spectrogram of the input, instead of performing it on the input speech features directly. Then, we detect adversarial examples by the recovered utterance, which is obtained by the iSTFT operation from the masked complex spectrogram. In this way, the hyperparameters are de-correlated with the dimensionality of the input features. For the hand-crafted mask problem, we attempt to obtain the mask matrix by a neural network instead of designing it manually, and replace the IBM matrix by either the ideal ratio masking (IRM) matrix or the approximate ideal binary masking (AIBM) matrix.

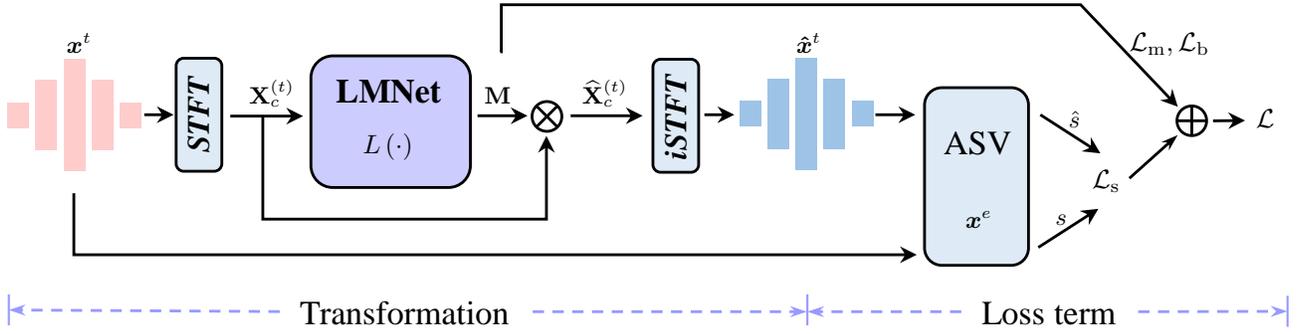
\begin{figure*}[t]
    \centering
    \includegraphics[width=0.95\textwidth]{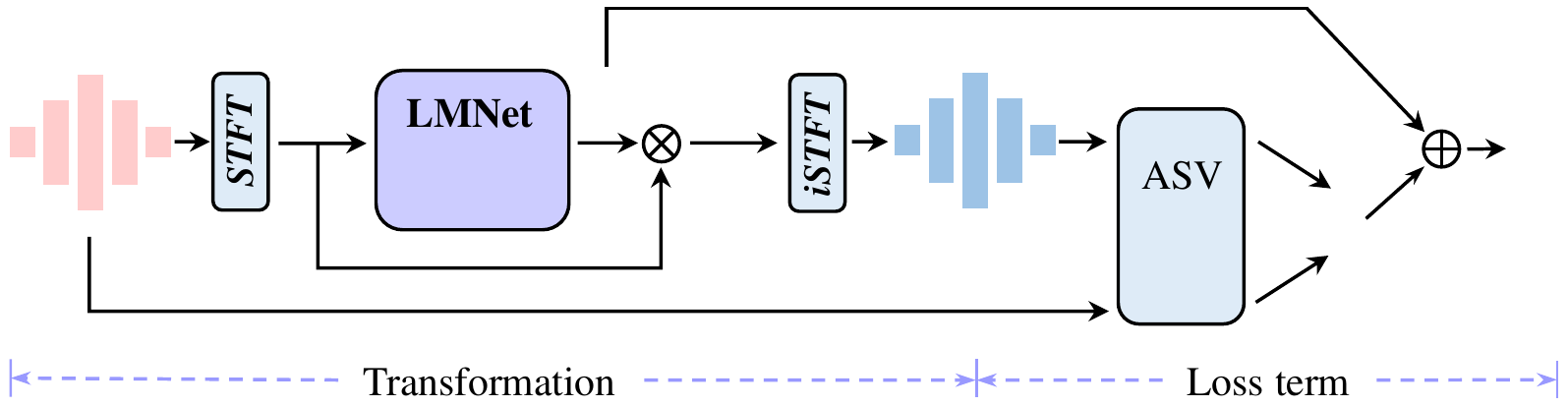}
    \put(-465,106){\small\bfseries\color{black}{$\bm{x}^t$}}
    \put(-396,87){\small\bfseries\color{black}{$\mathbf{X}_{c}^{\left(t\right)}$}}
    \put(-353,68){\normalsize\bfseries\color{black}{$L\left(\cdot\right)$}}
    \put(-307,87){\small\bfseries\color{black}{$\mathbf{M}$}}
    \put(-270,87){\small\bfseries\color{black}{$\mathbf{\widehat{X}}_{c}^{\left(t\right)}$}}
    \put(-189,106){\small\bfseries\color{black}{$\bm{\hat{x}}^t$}}
    \put(-125,40){\small\bfseries\color{black}{$\bm{x}^e$}}
    \put(-86,78){\small\bfseries\color{black}{$\hat{s}$}}
    \put(-91,41){\small\bfseries\color{black}{$s$}}
    \put(-77,55){\normalsize\bfseries\color{black}{$\mathcal{L}_{\rm s}$}}
    \put(-63,107){\normalsize\bfseries\color{black}{$\mathcal{L}_{\rm m},\mathcal{L}_{\rm b}$}}
    \put(-15,78){\normalsize\bfseries\color{black}{$\mathcal{L}$}}

    \caption{Training process of the Learnable Mask Detector (LMD). Given a genuine utterance $\bm{x}^t$, the loss function $\mathcal{L}$ in \eqref{equ:training_loss} takes the corresponding transformed utterance $\bm{\hat{x}}^{t}$ and the mask matrix $\mathbf{M}$ to train the learnable mask network (LMNet) $L\left(\cdot\right)$. The forward (black solid lines) and the gradients backward (red dashed lines) propagation process are shown. After the transformed utterance $\bm{\hat{x}}^{t}$ is obtained by the well-trained LMNet $L\left(\cdot\right)$, we begin the detection process in \figurename{\ref{fig:mcs_detect}}.}
    \label{fig:lmd_train}
\end{figure*}

\subsection{Framework}\label{subsec:framework}
The pipeline of the proposed method contains two steps: transformation and detection, as shown in \figurename{\ref{fig:mcs_detect}}. The proposed two methods, i.e. MCS and LMD, differ in the transformation process, and share the same detection module.
\subsubsection{Transformation}
Given an input test utterance $\bm{x}^t$, we first obtain its complex spectrogram $\mathbf{X}_{c}^{\left(t\right)}$ by the STFT operation,
\begin{equation}\label{equ:stft_transform}
    \mathbf{X}_{c}^{\left(t\right)} = g\left(\bm{x}^t; \phi \right),
\end{equation}
where $\mathbf{X}_{c}^{\left(t\right)}\in\mathbb{C}^{F\times T}$ with $F$ and $T$ representing the number of frequency bins and frames respectively, and $g\left(\cdot\, ; \phi\right)$ represents the STFT operator with parameters $\phi$, such as frame length, frame shift, and number of points of the fast fourier transform. Then we use $\mathbf{X}_{c}^{\left(t\right)}$ to  calculate a mask matrix $\mathbf{M}$ by either MCS or LMD, and perform the masking operation on the complex spectrogram $\mathbf{X}_{c}^{\left(t\right)}$ via:
\begin{equation}\label{equ:mask_dot_product}
    \mathbf{\widehat{X}}_{c}^{\left(t\right)}=\mathbf{M}\odot \mathbf{X}_{c}^{\left(t\right)},
\end{equation}
where $\mathbf{\widehat{X}}_{c}^{\left(t\right)}$ is the masked complex spectrogram, and $\odot$ denotes the element-wise product operator. Finally, the transformed utterance $\bm{\hat{x}}^t$ is obtained by:
\begin{equation}
    \bm{\hat{x}}^t = g^{-1}\left(\mathbf{\widehat{X}}_{c}^{\left(t\right)}; \phi\right),
\end{equation}
where $g^{-1}\left(\cdot\, ; \phi\right)$ is the iSTFT operator with the same parameters $\phi$ in \eqref{equ:stft_transform}.

\subsubsection{Detection}
 After the transformation process of MCS or LMD to $\bm{x}^t$, the transformed utterance $\bm{\hat{x}}^t$ is obtained. Then, two similarity scores are calculated by:
\begin{eqnarray}
    s =\mathbf{S} \Bigl(\bm{x}^{t}, \bm{x}^{e};\, \bm{\theta}, f\Bigr), \\
    \hat{s} =\mathbf{S} \Bigl(\bm{\hat{x}}^{t}, \bm{x}^{e};\, \bm{\theta}, f\Bigr).
\end{eqnarray}
Finally, the proposed method compares the score variation $\upsilon=|s-\hat{s}|$ with a detection threshold $\tau_{\mathrm{det}}$. When $\upsilon > \tau_{\mathrm{det}}$, the test utterance $\bm{x}^t$ is detected as an adversarial example, otherwise, it is considered as a genuine example.

\subsection{Masking Complex Spectrogram}\label{subsec:mcs}
MCS only uses the magnitude $\mathbf{X}_{m}^{\left(t\right)}$ of the complex spectrogram to calculate a mask matrix $\mathbf{M}\in \mathbb{R}^{F\times T}$. It masks complex spectrograms either at high frequencies (MCS-H) or using one-order difference (MCS-D).

MCS-H obtains the mask matrix by:
\begin{equation}\label{equ:mask_matrix_h}
    \mathbf{M} = \left[\begin{array}{l}
    \mathbf{1}_{\left(F-l\right) \times T}\\
    \mathbf{0}_{\;l \times T}
     \end{array}
     \right],
\end{equation}
where $l$ is the length of the masking, and the symbols $\mathbf{1}_{a \times b}$ (or $\mathbf{0}_{a \times b}$) denotes an all one (or zero) matrix with $a$ rows and $b$ columns.

MCS-D masks the time-frequency bins whose absolute values of the one-order difference along the frequency axis is smaller than a masking threshold $\xi$:
\begin{equation}
    \begin{array}{cc}
        &\mathbf{M}_{i,j} = \left\{\begin{array}{cl}
            1, & {}\mbox{ if } \Bigl| \mathbf{X}^{\left(t\right)}_{m\;\left(i+1,j\right)} - \mathbf{X}^{\left(t\right)}_{m\;\left(i,j\right)}\Bigr| > \xi \\
            \vspace{-5pt} \\
            0, & {} \mbox{ if } \Bigl| \mathbf{X}^{\left(t\right)}_{m\;\left(i+1,j\right)} - \mathbf{X}^{\left(t\right)}_{m\;\left(i,j\right)}\Bigr| \leq \xi
        \end{array}\right., \\
        \vspace{-2pt} \\
        &\forall\, i=1,2,\cdots,F-1, \quad \forall\, j=1,2,\cdots,T
    \end{array}
\end{equation}
where the subscripts $i$ and $j$ are the coordinates of the frequency axis and time axis, respectively. To make the mask matrix the same size as $\mathbf{X}_{c}^{\left(t\right)}$ in \eqref{equ:stft_transform}, we further concatenate an all-zero matrix $\mathbf{0}_{1 \times T}$ at the highest frequency, i.e., $\mathbf{M}_{F,j}=0$, $ \forall\, j=1,\cdots,T$.

\subsection{Learnable Mask Detector}\label{subsec:lmd}
As mentioned in Section~\ref{ssec:motivation}, the LMD detection method improves MCS by learning $\mathbf{M}$ automatically. It is worthy noting that (i) the learnable mask network (LMNet) of LMD only uses genuine examples for training, so it is insensitive to the parameters and types of adversarial examples, i.e. attacker-independent, and (ii) LMD obtains strong interpretability, since the training and detection phases of LMD are closely related. \figurename{\ref{fig:lmd_train}} illustrates the training process of LMD. We describe its transformation process and training loss for the masking generation as follows.

\subsubsection{Transformation Process}
As shown in the left part of \figurename{\ref{fig:lmd_train}}, there are two important differences between the transformation of LMD and MCS. First, the complex spectrogram feature are explicitly divided into real and imaginary parts, as $\mathbf{X}_{c}^{\left(t\right)}\in\mathbb{R}^{F\times T\times 2}$. Second, the mask matrix $\mathbf{M}$ with the same size of $\mathbf{X}_{c}^{\left(t\right)}$ is obtained by the well-trained LMNet $L\left(\cdot\right)$.

\begin{algorithm}[t!]
    \small
    \caption{\small Training procedure of LMD.}
    \label{alg:lmnet_training_process}
    \LinesNumbered
    \KwIn{The training data $\mathcal{D}^{t}$, the validation data $\mathcal{D}^{v}$, and the defensive ASV model $\mathbf{S} \left(\cdot\, ;\, \bm{\theta},f\right)$.}
    \KwOut{The well-trained LMNet $L\left(\cdot\right)$ with parameters $\Psi^{*}$.}

    \BlankLine
    \BlankLine
    Initialize the hyperparameters $m$, $\lambda_{\text{s}}$, and $\lambda_{\text{b}}$\;
    \BlankLine

    \While{the number of training iterations}{
        \BlankLine
        $\mathbf{X}^t \leftarrow$ minibatch of $q$ samples from $\mathcal{D}^{t}$\;
        $\mathbf{X}^e \leftarrow$ randomly sampling the utterances of the same speaker with $\mathbf{X}^t$ from $\mathcal{D}^{t}$\;

        \BlankLine
        $\mathbf{M}, \mathbf{\widehat{X}}^t \leftarrow$ Propagate the minibatch data $\mathbf{X}^t$ forward the LMNet $L\left(\cdot\right)$ as shown in \figurename{\ref{fig:lmd_train}}\;

        \BlankLine
        Compute the loss function \eqref{equ:training_loss}, as Loss $\leftarrow$ $\frac{1}{q}\sum\limits_{i}^{q}\mathcal{L}\left(\mathbf{X}^t, \mathbf{X}^e, \mathbf{M}, \mathbf{\widehat{X}}^t; m, \lambda_{\text{s}},\lambda_{\text{b}}, \mathbf{S}\right)$\;
        Minimize the loss function to update $L\left(\cdot\right)$\;

        \BlankLine
        \If{reach the validation iteration interval}{
            Compute the loss \eqref{equ:training_loss} from the validation data $\mathcal{D}^{v}$, denoted as validation loss, and update the best parameters $\Psi^{*}$ based on the validation loss\;
        }
    }
    \BlankLine
\end{algorithm}

\subsubsection{Training Loss}
The right part of \figurename{\ref{fig:lmd_train}} describes the loss function of LMD. The design of the training loss is based on the assumption that adversarial perturbations are uniformly distributed in the feature space (e.g. the complex spectrograms), which makes us believe that the more the time-frequency bins are masked, the more likely the adversarial examples are to fail. However, when more time-frequency bins are masked, the discriminability of the ASV to the genuine examples decreases as well.

To address the above contradictory effects simultaneously, we expect to mask out as much as possible the time-frequency bins that contain little speaker information. Three loss terms are designed for this purpose:

The first loss term $\mathcal{L}_{\text{s}}$ is the score variation, which measure the amount of speaker information contained in the masked time-frequency bins:
\begin{equation}\label{equ:loss_term_variation}
    \mathcal{L}_{\text{s}} = \max\Bigl(0, |s - \hat{s}| - m\Bigr),
\end{equation}
where $m$ is a margin\footnote{Unless specified otherwise, the margin $m$ is set to $0.05$ in our LMD.} of the hinge-loss, which is used to quantify the magnitude of the score variation, and the score $s$ is the cosine similarity of the speaker embeddings of the test utterance $\bm{x}^t$ and the enrollment utterance $\bm{x}^e$, and $\hat{s}$ is the cosine similarity of the speaker embeddings of the transformed utterance $\bm{\hat{x}}^t$ and $\bm{x}^e$.

The second loss term $\mathcal{L}_{\text{b}}$ is the binary penalty for an AIBM matrix:
\begin{equation}\label{equ:loss_term_ibm}
    \mathcal{L}_{\text{b}} = \bigl\|\mathbf{M} \odot \left(\mathbf{1} - \mathbf{M}\right)\bigr\|_{2}^{2},
\end{equation}
where the symbol $\mathbf{1}$ represents an all one matrix of the same shape as $\mathbf{M}$. The binary penalty loss term will force the elements of the mask matrix to either converge to 0 or converge to 1, i.e., an AIBM matrix will be achieved.

The third loss term $\mathcal{L}_{\text{m}}$ is an $L_1$ norm of the mask matrix, which represents the severity of the masking operation:
\begin{equation}\label{equ:loss_term_mask}
   \mathcal{L}_{\text{m}}=\left\|\mathbf{M}\right\|_{1}.
\end{equation}

Finally, we propose to train LMNet by minimize the following loss function:
\begin{equation}\label{equ:training_loss}
    \mathcal{L} = \mathcal{L}_{\text{m}} + \lambda_{\text{s}}  \mathcal{L}_{\text{s}} +  \lambda_{\text{b}}  \mathcal{L}_{\text{b}},
\end{equation}
where $\lambda_{\text{s}}$ and $\lambda_{\text{b}}$ are the hyperparameters used to balance the three loss terms. See Algorithm~\ref{alg:lmnet_training_process} for the complete training process of LMD.

\section{Experimental Setup} \label{sec:exp_setup}

\subsection{Datasets} \label{ssec:datasets}
All of our experiments were conducted on the VoxCeleb dataset \cite{nagrani2017voxceleb}, which contains over one million utterances from 7,363 speakers of different ethnicities, accents, professions, and ages. The VoxCeleb datasets are automatically collected from interview videos uploaded to YouTube, and the speech segments were contaminated with real-world noise. The two victim ASV models were trained on the development set of VoxCeleb2 \cite{chung2018voxceleb2} and evaluated on the cleaned up version of the original verification test list, i.e. {\Courier VoxCeleb1-test}, which consists of 37,611 trials from 40 speakers.

Without loss of generality, we randomly selected 1,000 trials from the original test list, denoted as the attack list {\Courier VoxCeleb1-attack}, to generate the adversarial examples. The randomly selected attack list include 500 target trials and 500 non-target trials. We also constructed an evaluation list {\Courier VoxCeleb1-eval} based on the attack list to evaluate the performance of attackers and detectors. The enrollment utterances of the evaluation list were randomly replaced with utterances of the same speaker in the test set of VoxCeleb1, but all utterances in the attack list were excluded.

Note that our proposed methods, MCS and LMD, were trained on the {\Courier VoxCeleb1-dev} dataset, which do not have overlapped speakers with the {\Courier VoxCeleb1-test} list. Moreover, {\Courier VoxCeleb1-dev} was divided into a training subset $\mathcal{D}^{t}$ and a validation subset $\mathcal{D}^{v}$ with a ratio of 19:1.

\subsection{Experimental Settings}
\subsubsection{Victim ASV Systems}
Different ASV models are characterized by different network structures, pooling strategies and training objectives. In this study, we used two ASV models as the victim. The first one is the ECAPA\_TDNN\footnote{https://github.com/wenet-e2e/wespeaker} \cite{desplanques2020ecapa} with a classification-based loss (AAM-Softmax \cite{liu2019large}) and the attentive statistical pooling. The second one is the Fast-ResNet34\footnote{https://github.com/clovaai/voxceleb\_trainer} with a verification-based loss (Angular Prototypical \cite{chung2020defence}) and attentive average pooling. They adopted the same acoustic feature extractor: a hamming window of width 25ms with a step size of 10ms was used to partition speech signals into frames, and a 80-dimensional LogFBank followed by cepstral mean and variance normalization (CMVN) were extracted as the acoustic features. Online data augmentation, such as perturbing speed, superimposed disturbance, and simulating reverberation were adopted in the training process. In addition, they all used cosine similarity as the back-end scoring. The system decision threshold $\eta$ is picked to be the threshold corresponding to the EER on the {\Courier VoxCeleb1-test}.

\begin{algorithm}[t!]
    \small
    \caption{\small Method for searching the hyperparameters of MCS.}
    \label{alg:mcs_hyperparameter_search}
    \LinesNumbered
    \KwIn{The training data $\mathcal{D}^{t}$.}
    \KwOut{The optimal hyperparameter $p$.}

    \BlankLine
    \BlankLine
    Initialize $\lambda_{\text{s}}=10,\lambda_{\text{b}}=0,m=0.1$ in \eqref{equ:training_loss}, $p_{\text{lower}}=0$, $p_{\text{upper}} = 257$ for MCS-H and $p_{\text{upper}} = 10^5$ for MCS-D\;
    \BlankLine

    \While{the maximum number of search has not been reached {\rm \textbf{or}} $\lvert p_{\text{\rm upper}} - p_{\text{\rm lower}} \rvert >= 1$}{
        \BlankLine
        Divide the interval $\left[p_{\text{lower}}, p_{\text{upper}}\right]$ into four equal parts, and obtains three parameters $p_1, p_2, p_3$ with an ascending order\;

        \BlankLine
        Load a minibatch data from $\mathcal{D}^{t}$, and randomly select utterances of the same speaker as the enrollment\;

        \BlankLine
        Calculate the loss value corresponding to the three obtained parameters by \eqref{equ:training_loss}, denoted as $L_1, L_2, L_3$. Note that the mask matrix $\mathbf{M}$ and transformed utterance $\bm{\hat{x}}^t$ were crafted by MCS-H or MCS-D\;

        \BlankLine
        \uIf{$L_1$ is the smallest loss of the three}{
            $p_{\text{upper}} \leftarrow p_2$
        }
        \uElseIf{$L_3$ is the smallest loss of the three}{
            $p_{\text{lower}} \leftarrow p_2$
        }\Else{
            $p_{\text{lower}} \leftarrow p_1, p_{\text{upper}} \leftarrow p_3$
        }
    }
    \BlankLine
    \tcp{Note the results are rounded.}
    \KwResult{$\left(p_{\text{lower}} + p_{\text{upper}}\right) / 2$}
\end{algorithm}

\begin{table}[t!]
    \centering
    \caption{Statistical results of the searched hyperparameters of MCS-H and MCS-D over ten independent runs of Algorithm \ref{alg:mcs_hyperparameter_search} on the {\Courier VoxCeleb1-dev} dataset. The means of the hyperparameters were adopted in other experiments.}
    \begin{tabular}{lcc}
        \toprule
        $\text{mean}\pm \text{std}$ & \multicolumn{1}{c}{\makecell{ECAPA\_TDNN + \\ AAM-Softmax}} & \multicolumn{1}{c}{\makecell{Fast-ResNet34 + \\ Angular Prototypical}} \\
        \midrule
        MCS-H $\rightarrow l$ & $79 \pm 2$    & $120 \pm 10$ \\
        MCS-D $\rightarrow \xi$& $643 \pm 71$   & $1164 \pm 285$ \\
        \bottomrule
    \end{tabular}
    \label{tab:mcs_hyperparameter_results}
\end{table}

\subsubsection{Attackers} \label{ssec:attacker}
We generated adversarial examples for three attackers based on the attack list {\Courier VoxCeleb1-attack}. For the BIM and PGD attackers, with the step size $\alpha=1$ and $\alpha=300$ fixed respectively, we generated adversarial examples for each value of the maximum iterations $N$, and constructed the \textit{adversarial trial set} $\mathcal{A}_{i}$, where $i=1,2,\cdots,6$, and $N=5,10,20,50,100,200$. For the CW attacker, with the maximum number of binary search and iterations, $N_{\text{bs}}=9$ and $N=100$, respectively, we also constructed adversarial trial set $\mathcal{A}_{i}$ for each value of the confidence $\kappa$, where $\kappa=0.0, 0.1, 0.2, 0.3, 0.4, 0.5$. We denotes the mixture set of the adversarial trials, i.e. \textit{adversarial mixture set}, crafted by the BIM attacker as $\mathcal{A}_{\text{\tiny BIM}}=\{\mathcal{A}_i \mid i=1,2,\cdots,6\}$. For the PGD and CW attackers, $\mathcal{A}_{\text{\tiny PGD}}$ and $\mathcal{A}_{\text{\tiny CW}}$ were defined similarly with $\mathcal{A}_{\text{\tiny BIM}}$. In addition, the corresponding \textit{genuine trial set} $\mathcal{G}_i$ was constructed by adding the Gaussian white-noise to the original clean utterances in the attack list, which aims to obtain the same SNR as the adversarial utterances in $\mathcal{A}_i$. The black-box attacker employed in this paper is the transfer-based cross-model attacker, i.e., the adversarial example generated by one substitute ASV is used to attack another victim ASV.

\subsubsection{Defenders}
The baseline detectors are the Vocoder, GL-mel, and GL-lin respectively, all of which followed the settings in \cite{wu2022adversarial}. They also utilize the score variation for detection, and the difference is that the phase reconstruction transformation are performed on the input utterances by vocoders, such as the PWG model. The settings of the masking length $l$ and masking threshold $\xi$ for the proposed MCS-H and MCS-D are shown in \tablename{\ref{tab:mcs_hyperparameter_results}}, which were determined by Algorithm~\ref{alg:mcs_hyperparameter_search}. The LMNet of the proposed LMD, which uses the network structure of DCCRN \cite{hu2020dccrn}, aims to obtain a mask matrix with high generalization by the complex convolution. The complex spectrogram was extracted as the input feature by a hanning window of width 25ms plus a step size of 10ms and the convolutional STFT. The batch size was set to 32 and the length of each audio clip was set to 500 frames. The Adam optimizer with an initial learning rate of 0.002 was used to train the LMNet $L\left(\cdot\right)$ guided by the loss in \eqref{equ:training_loss}, where the hyperparameter $\lambda_{\text{s}}$ was set to 1. The hyperparameter $\lambda_{\text{b}}$ in \eqref{equ:training_loss} controls the type of the mask matrix\footnote{$\lambda_{\text{b}}=15$ indicates the LMD-AIBM, and $\lambda_{\text{b}}=0$ indicates the LMD-IRM.}. The learning rate was decayed by 0.9 times for every 1,000 steps. A total of 30K iterations were trained, and the optimal model was selected based on the validation data $\mathcal{D}^{v}$ with a validation interval of 1,000 steps.

\begin{figure}
    \centering
    \includegraphics[width=0.95\linewidth]{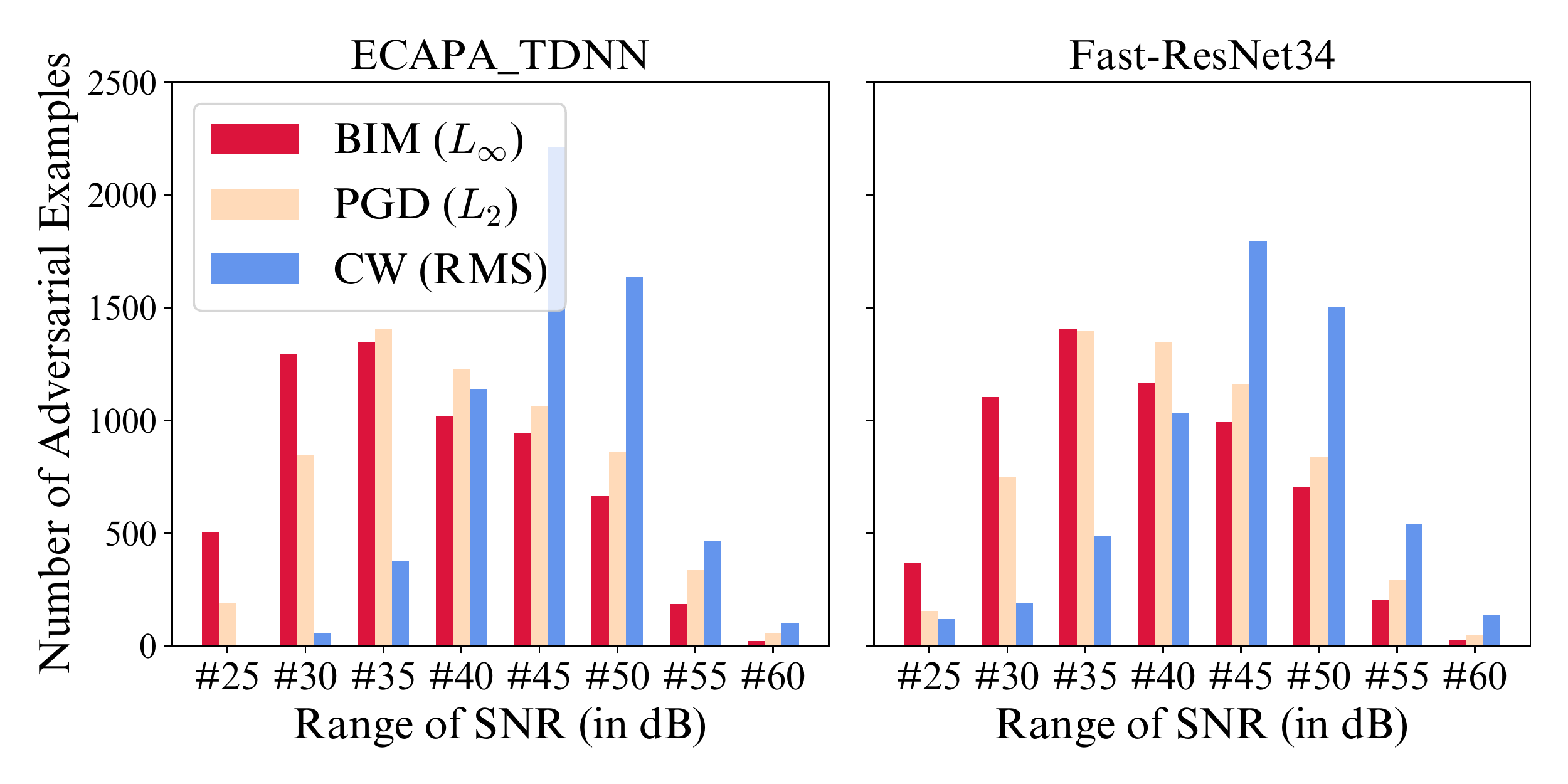}
    \caption{Statistical results of the number of adversarial examples in a SNR range. The ECAPA\_TDNN and Fast-ResNet34 act as the victim ASV. The symbol ``\#$n$'' means the range of ``$\left[n, n+5\right)$''.}
    \label{fig:attackers_aenumber}
\end{figure}

\begin{figure*}
    \centering
    \subfigure[White-box attacks on ECAPA\_TDNN]{
        \label{sfig:white-ecapatdnn}
        \includegraphics[width=0.3\linewidth]{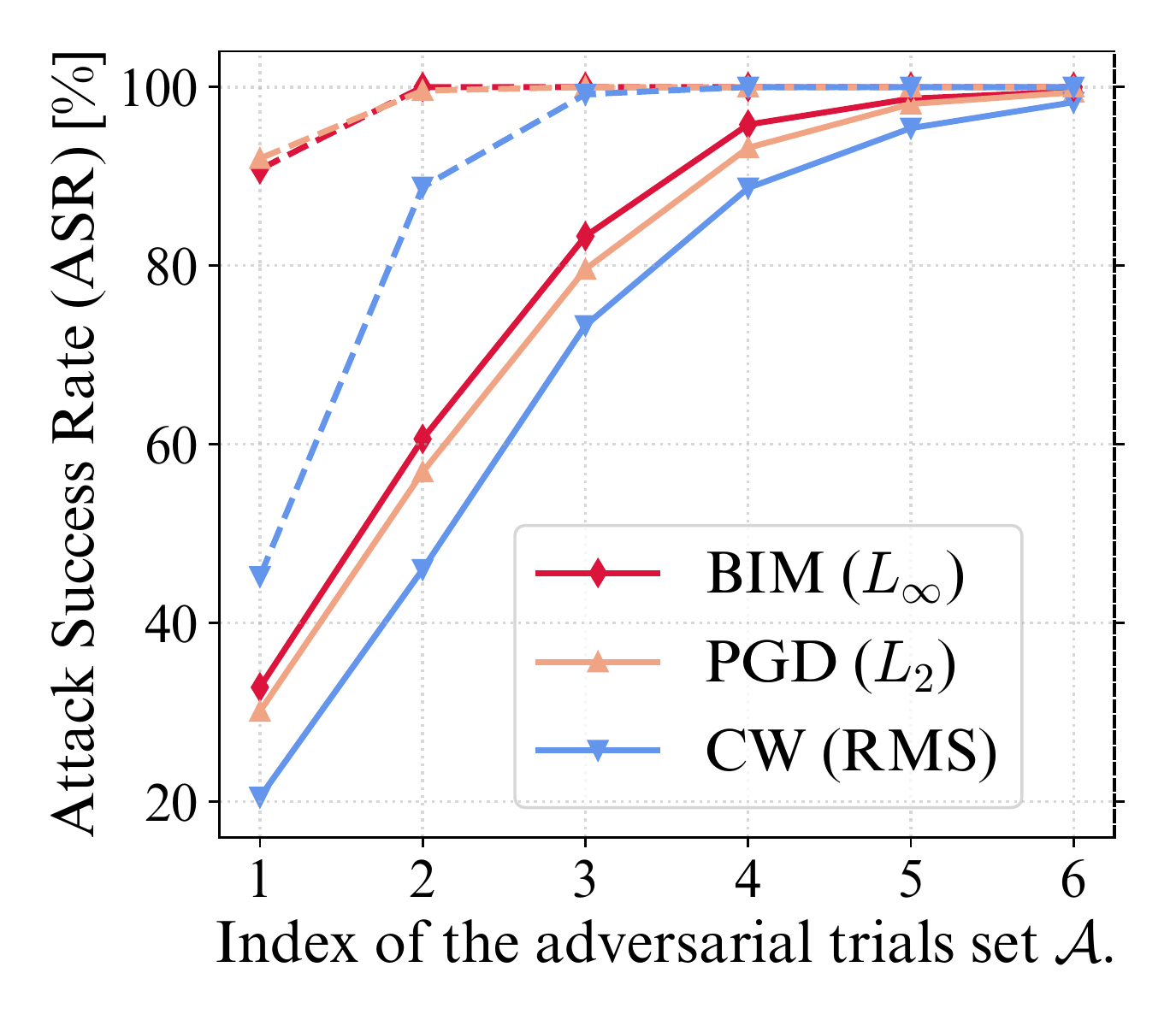}
    }
    \hspace{-18pt}
    \subfigure[White-box attacks on Fast-ResNet34]{
        \label{sfig:white-resnetap}
        \includegraphics[width=0.3\linewidth]{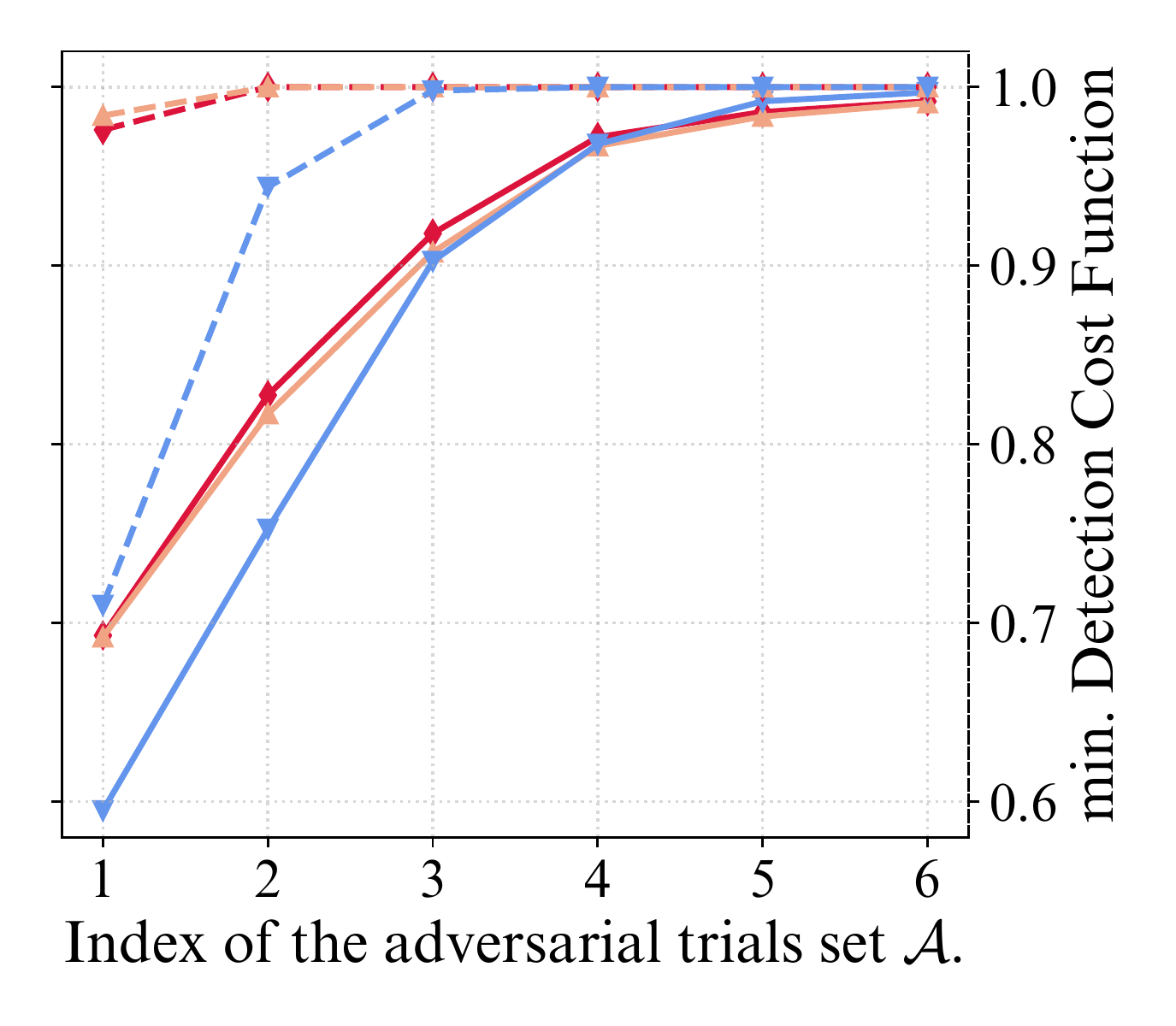}
    }
    \hspace{-18pt}
    \subfigure[Adversarial examples for ECAPA\_TDNN]{
        \label{sfig:aesnr-ecapatdnn}
        \includegraphics[width=0.3\linewidth]{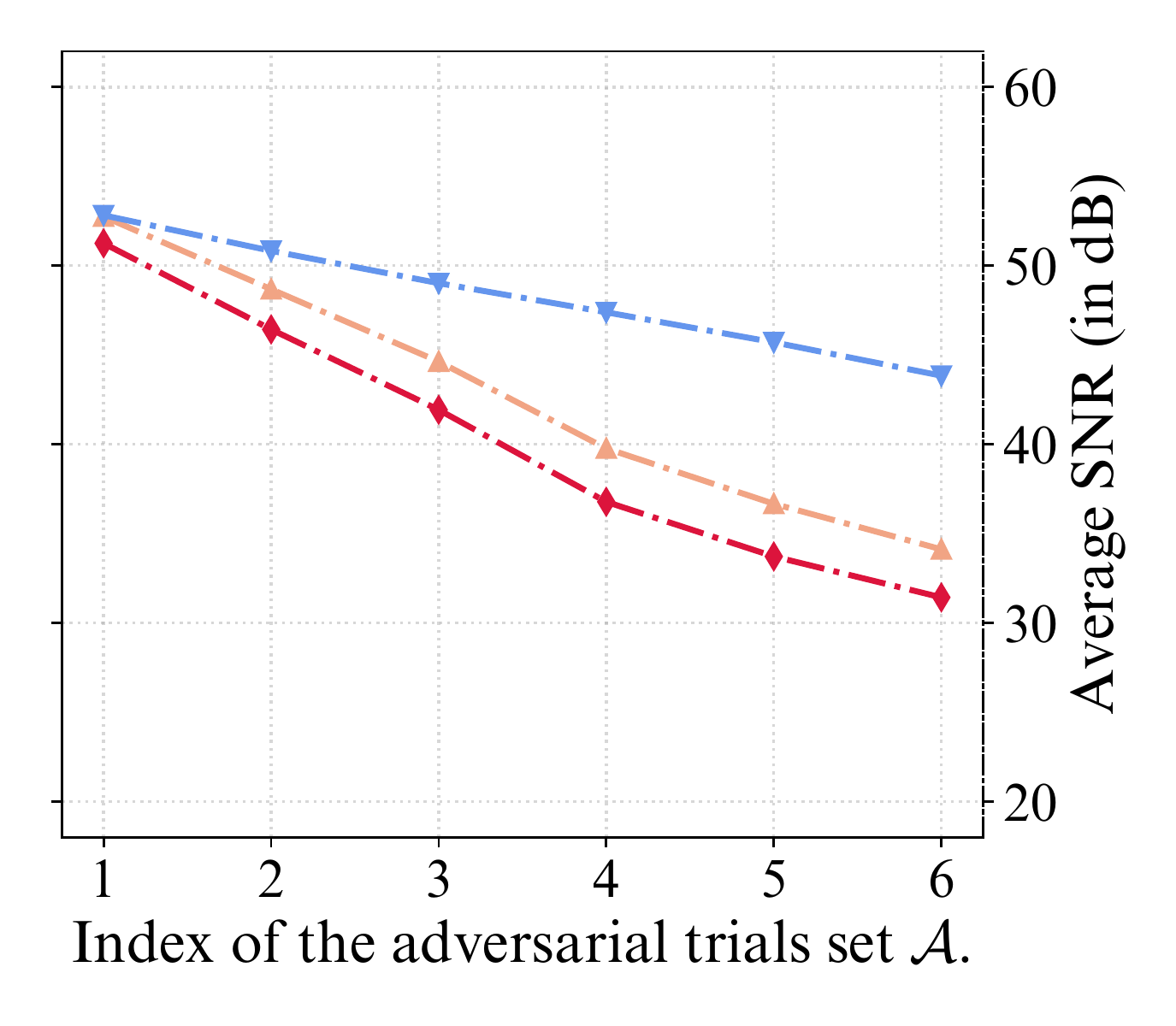}
    }

    \subfigure[Black-box attacks on ECAPA\_TDNN]{
        \label{sfig:black-ecapatdnn}
        \includegraphics[width=0.3\linewidth]{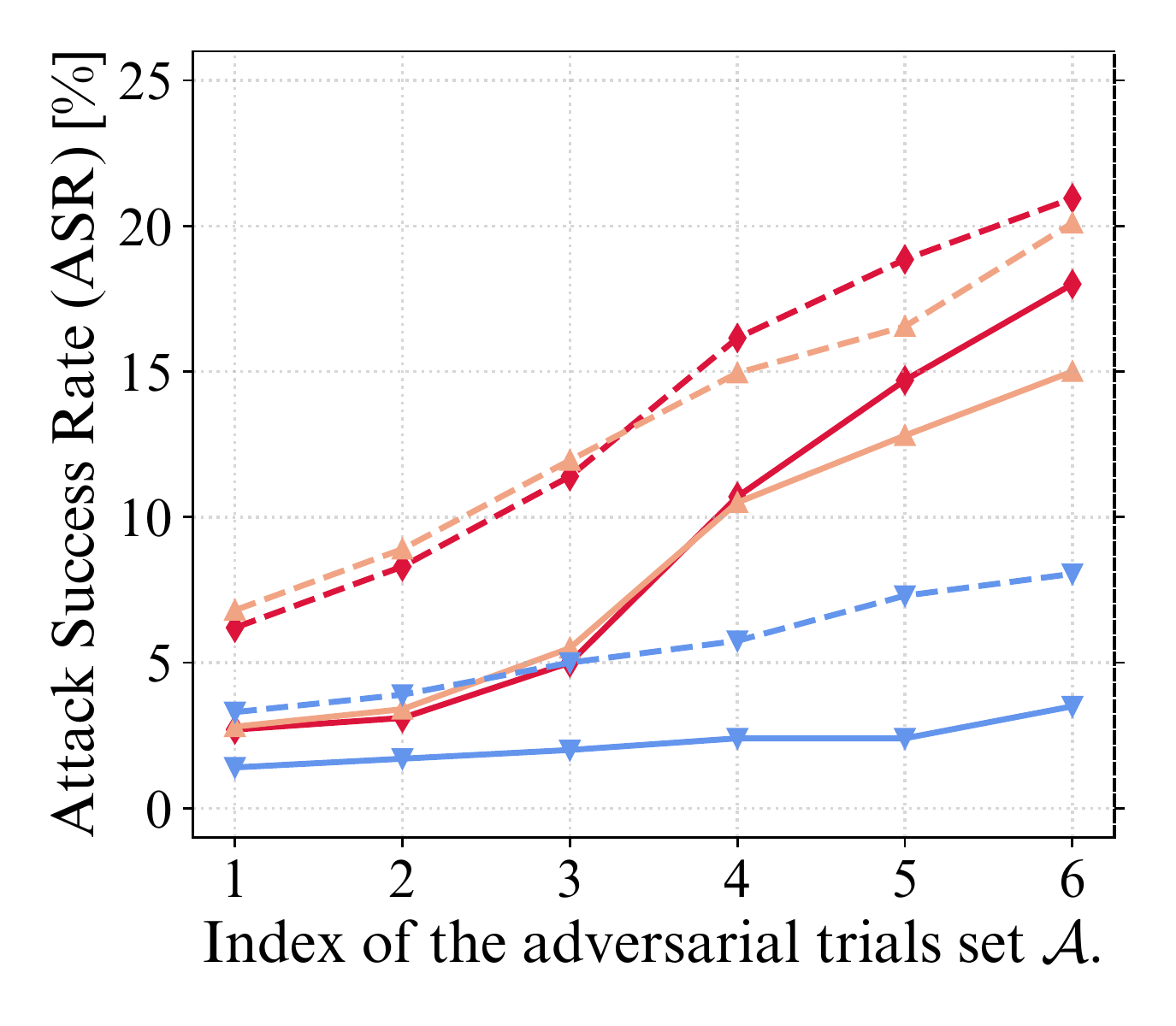}
    }
    \hspace{-18pt}
    \subfigure[Black-box attacks on Fast-ResNet34]{
        \label{sfig:black-resnetap}
        \includegraphics[width=0.3\linewidth]{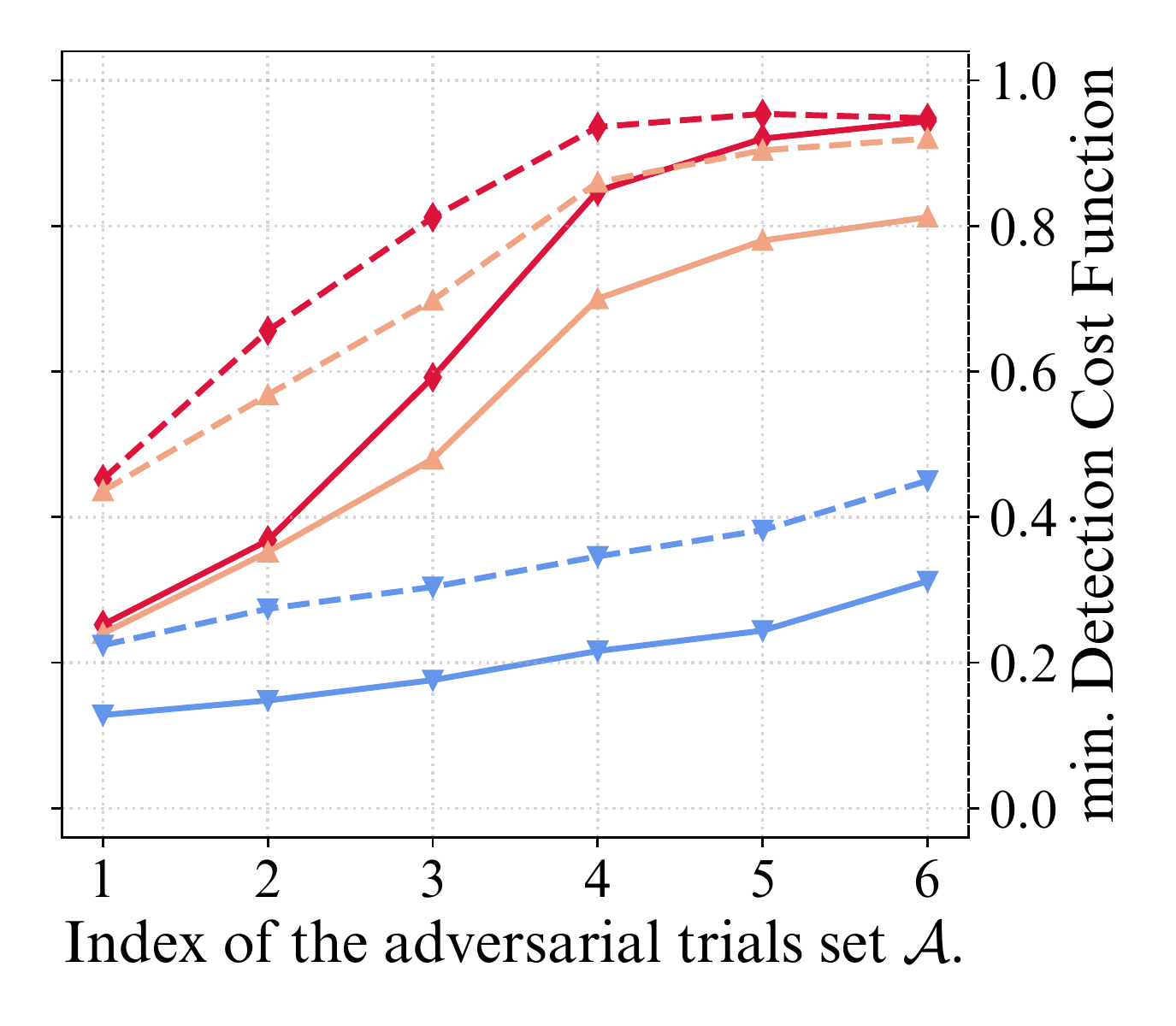}
    }
    \hspace{-18pt}
    \subfigure[Adversarial examples for Fast-ResNet34]{
        \label{sfig:aesnr-resnetap}
        \includegraphics[width=0.3\linewidth]{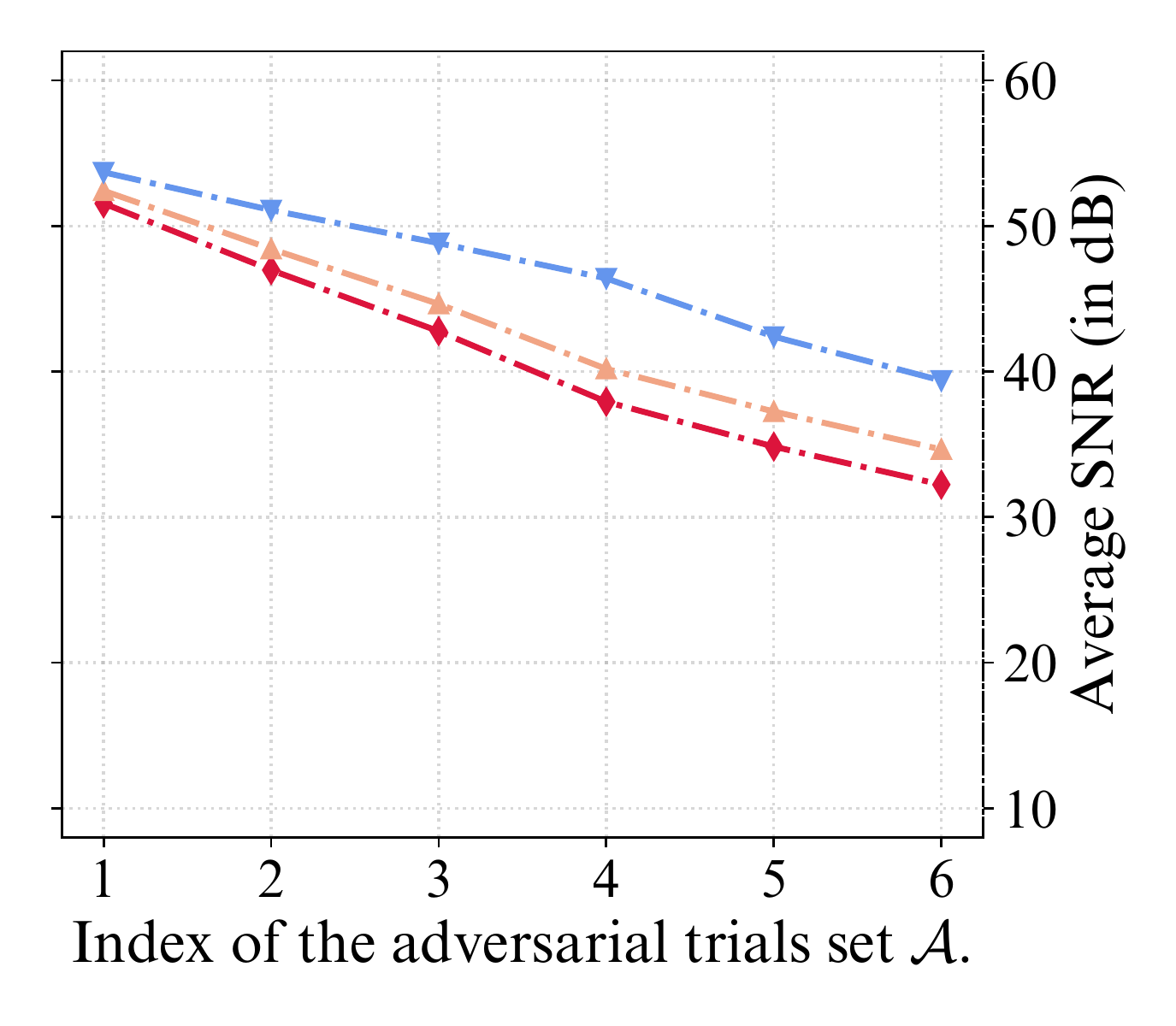}
    }

    \caption{Attack performance of the three attackers: BIM, PGD and CW in terms of ASR, minDCF, and mean SNR, where \textbf{ASR is described in solid line, the minDCF with $\bm{p=0.01}$ is described in dashed line, and the mean SNR is described in dotted line}. The captions of the subfigures ``(a), (b), (d), (e)'' are concise. For example, ``Balck-box attacks on ECAPA\_TDNN'' means that the victim and substitute ASV models are ECAPA\_TDNN and Fast-ResNet34, respectively. The subfigures ``(c)'' and ``(f)'' depict the average SNR of the adversarial examples. Note that, the EER of the ECAPA\_TDNN ASV model with the AAM-Softmax loss on the test list {\Courier VoxCeleb1-test} is 1.25\%; the EER of the Fast-ResNet34 ASV model with the Angular Prototypical is 1.97\%.
    }
    \label{fig:attackers_perf}
\end{figure*}

\subsection{Evaluation Metrics}
To evaluate the harmfulness of the attackers, we employ the attack success rate (ASR), normalized minimum detection cost function (minDCF) of the victim ASV with $p=0.01$ and $C_{\text{miss}}=C_{\text{fa}}=1$ \cite{nagrani2020voxsrc}, and SNR, as the evaluation metrics.

To evaluate the performance of the detectors, we adopt EER and the detection success rate (DSR) with different given false acceptance rate (FAR), as the evaluation metrics.

Before introducing the evaluation metrics, we first define the score variation set for the genuine trial set and adversarial trial set, respectively. For the genuine trial set $\mathcal{G}=\left\{\left(\bm{x}^t_{i},\bm{x}^e_{i}\right) \mid i=1,2,\cdots,I\right\}$ defined in Section~\ref{ssec:attacker}, a score variation set $\mathcal{V}_{\mathrm{gen}}$ after the masking operation can be obtained by:
\begin{equation}\label{equ:score_variation}
    \upsilon_{i} = \Bigl| \mathbf{S} \Bigl(\bm{x}^{t}_i, \bm{x}^{e}_i;\, \bm{\theta}, f\Bigr) - \mathbf{S} \Bigl(\bm{\hat{x}}^{t}_i, \bm{x}^{e}_i;\, \bm{\theta}, f\Bigr) \Bigr|
\end{equation}
where $\upsilon_{i}\in\mathcal{V}_{\mathrm{gen}}$ with $i=1,2,\cdots,I$, and $\bm{\hat{x}}^{t}_i$ represents that the test utterance $\bm{x}^t_{i}$ is transformed by our mask-based detection methods. For the adversarial trial set $\mathcal{A}=\left\{\left(\bm{\tilde{x}} ^t_{i},\bm{x}^e_{i}\right) \mid i=1,2,\cdots,I\right\}$, its score variation set $\mathcal{V}_{\mathrm{adv}}$ is also calculated by \eqref{equ:score_variation}, except that $\bm{x}^t_{i}$ and $\bm{\hat{x}}^{t}_i$ are replaced by the corresponding adversarial example $\bm{\tilde{x}} ^t_{i}$ and the transformed adversarial example, respectively.

Then the evaluation metric EER is defined by:
\begin{equation}\label{equ:eer_define}
    \text{EER}_{\mathrm{det}} = \text{FAR}_{\mathrm{det}}\left(\tau_{\mathrm{eer}}\right) = \text{FRR}_{\mathrm{det}}\left(\tau_{\mathrm{eer}}\right),
\end{equation}
where
\begin{equation}
    \text{FAR}_{\mathrm{det}}\left(\tau\right) = \frac{\left|\left\{\upsilon_{i}>\tau \mid \upsilon_{i}\in \mathcal{V}_{\mathrm{gen}}\right\}\right|}{\left| \mathcal{V}_{\mathrm{gen}} \right|},
\end{equation}
\begin{equation}
    \text{FRR}_{\mathrm{det}}\left(\tau\right) = \frac{\left|\left\{\upsilon_{i}\leq\tau \mid \upsilon_{i}\in \mathcal{V}_{\mathrm{adv}}\right\}\right|}{\left| \mathcal{V}_{\mathrm{adv}} \right|},
\end{equation}
are the FAR and the false rejection rate (FRR), respectively, of the detector given a threshold $\tau$, $\left| \mathcal{S}\right|$ represents the number of the elements in the set $\mathcal{S}$. After manually given a tolerable FAR of detection, denoted as $\text{FAR}_{\mathrm{given}}$, we define the evaluation metric DSR as:
\begin{equation}\label{equ:dsr_define}
    \text{DSR} = \frac{\left|\left\{\upsilon_{i}>\tau_{\mathrm{det}} \mid \upsilon_{i}\in \mathcal{V}_{\mathrm{adv}}\right\}\right|}{\left| \mathcal{V}_{\mathrm{adv}} \right|},
\end{equation}
where
\begin{equation}\label{equ:det_threshold_define}
    \tau_{\mathrm{det}} = \operatornamewithlimits{argmin}_{\tau} \,\bigl| \text{FAR}_{\mathrm{det}}\left(\tau\right)-\text{FAR}_{\mathrm{given}} \bigr| ,
\end{equation}
is the detection threshold for $\text{FAR}_{\mathrm{given}}$.
We also evaluated the DSR of detectors under the adversarial mixture sets, i.e. $\mathcal{A}_{\text{\tiny BIM}}$, $\mathcal{A}_{\text{\tiny PGD}}$ and $\mathcal{A}_{\text{\tiny CW}}$.

Finally, the detection EER is defined over a given SNR budget, as in \cite{villalba2020x}. Specifically, we assume an evaluator function $\mathbf{E}\left(\mathcal{A}, \mathcal{G}\right)$ that computes the detector EER given the adversarial trial set $\mathcal{A}$ and genuine trial set $\mathcal{G}$ with $I$ trials. We assume that $\bm{p}_{\text{adv}} = [p_{\text{adv},1},\ldots,p_{\text{adv},I}]^T$ and $\bm{p}_{\text{gen}}= [p_{\text{gen},1},\ldots,p_{\text{gen},I}]^T$ are vectors describing the SNRs of the corresponding trail sets respectively. Then, for each value of the SNR budget $b$ that we want to evaluate, we obtain an adversarial trial set $\mathcal{A}\left(b\right)$ and a genuine trails set  $\mathcal{G}\left(b\right)$ by:
\begin{equation}
    t_i = \left\{\begin{array}{cc}
        \left(\bm{\tilde{x}} ^t_{i},\bm{x}^e_{i}\right), &\text{if }{p}_{\text{adv},i} \geq b \text{ or }{p}_{\text{gen},i} \geq b \\
        \varnothing, &\text{otherwise}
    \end{array}\right.,
\end{equation}
where $t_i \in \mathcal{A}\left(b\right)$ with $ i=1,2,\cdots,I$, and $\mathcal{G}\left(b\right)$ is composed of the corresponding trials in $\mathcal{G}$. The detector EER for budget $b$ is obtained by evaluating $\mathbf{E}\bigl(\mathcal{A}\left(b\right), \mathcal{G}\left(b\right)\bigr)$.

\begin{table*}[t]
    \centering
    \caption{Detection EER of the detectors against three attackers in the white-box attack scenario on the two ASVs.}
    \scalebox{0.85}{\begin{tabular}{p{60pt}<{\centering}c
                                    *{6}{p{12pt}<{\centering}}   p{2pt}<{\centering}
                                    *{6}{p{12pt}<{\centering}}   p{2pt}<{\centering}
                                    *{6}{p{12pt}<{\centering}}}
    \toprule[1.3pt]
    EER (\%) $\searrow $ & Attacker $\rightarrow$ & \multicolumn{6}{c}{BIM ($L_{\infty}$, $\alpha=1$)} &  & \multicolumn{6}{c}{PGD ($L_2$, $\alpha=300$)} &  & \multicolumn{6}{c}{CW (RMS, $N=100$)} \\
    \specialrule{0em}{3pt}{3pt}
    Victim Model $\downarrow$ & $N / \kappa\; \rightarrow$ & 5     & 10    & 20    & 50    & 100   & 200 &   & 5     & 10    & 20    & 50    & 100   & 200 &  & 0     & 0.1   & 0.2   & 0.3   & 0.4   & 0.5 \\
    \midrule[0.8pt]

    \multirow{7}{*}{\makecell{ECAPA\_TDNN + \\ AAM-Softmax}}
    & Vocoder   & 11.50  & 4.30  & \textbf{1.70}  & 1.30  & 2.00  & 2.00  & & 12.50  & 4.70  & \textbf{2.10 } & \textbf{1.10 } & 1.10  & 1.50  & & 11.60  & 5.20  & 2.70  & 1.80  & \textbf{0.50 } & \textbf{0.30 } \\
    & GL-mel    & 28.00  & 14.00 & 5.60  & 2.50  & 2.20  & 2.80  & & 28.60  & 14.60  & 6.80  & 2.90  & 2.10  & 2.00  & & 26.40  & 16.00  & 9.00  & 4.50  & 2.70  & 1.40  \\
    & GL-lin    & 20.80  & 9.80  & 3.90  & 2.20  & 2.10  & 2.90  & & 21.60  & 10.30  & 5.10  & 2.30  & 2.10  & 2.50  & & 19.20  & 10.50  & 5.00  & 3.00  & 1.80  & 1.50  \\
    & MCS-H     & 24.70  & 14.00 & 8.10  & 5.80  & 5.90  & 6.70  & & 26.30  & 15.30  & 9.00  & 5.80  & 5.60  & 6.00  & & 23.40  & 15.20  & 9.90  & 6.70  & 4.10  & 3.00  \\
    & MCS-D     & 15.50  & 8.40  & 5.20  & 2.90  & 3.20  & 2.70  & & 15.30  & 8.50   & 4.90  & 3.10  & 2.80  & 2.40  & & 15.50  & 9.10   & 5.40  & 3.50  & 2.60  & 1.90  \\
    & LMD ($\lambda_{b}=15$) & 16.10  & 6.70  & 2.30 & \textbf{0.80 } & \textbf{0.90 } & \textbf{1.20 } & & 15.90  & 6.70  & 2.50  & \textbf{1.10 } & \textbf{0.80 } & \textbf{1.10 } & & 14.50  & 7.50  & 3.80  & 1.50  & 0.90  & \textbf{0.30 } \\
    & LMD\; ($\lambda_{b}=0$)\;  & \textbf{7.70 } & \textbf{3.60 } & 3.00  & 4.10  & 4.60  & 6.70  & & \textbf{7.90 } & \textbf{4.40 } & 3.80  & 4.20  & 6.10  & 7.00  & & \textbf{9.10 } & \textbf{4.20 } & \textbf{2.00 } & \textbf{1.20 } & 0.90  & 1.10  \\
    \midrule[0.8pt]

    \multirow{7}{*}{\makecell{Fast-ResNet34 + \\ Angular Prototypical}}
    & Vocoder   & 12.70  & \textbf{5.00 } & \textbf{2.20 } & 1.70  & 2.00  & \textbf{2.00 } & & 12.00  & \textbf{5.10 } & \textbf{1.90 } & 1.80  & 1.70  & \textbf{1.80 } & & 17.70  & 8.60  & \textbf{2.90 } & \textbf{1.20 } & \textbf{0.80 } & 1.60  \\
    & GL-mel    & 23.50  & 11.50  & 5.20  & 2.40  & 3.10  & 3.40  & & 24.10  & 10.60 & 5.20  & 2.90  & 3.20  & 3.90  & & 28.50  & 17.30  & 9.20  & 4.00  & 2.30  & 3.20   \\
    & GL-lin    & 16.50  & 7.60   & 3.50  & 2.60  & 2.90  & 4.10  & & 15.60  & 6.80  & 3.60  & 2.90  & 3.80  & 4.10  & & 22.30  & 11.90  & 5.30  & 2.70  & 2.30  & 5.10   \\
    & MCS-H     & 30.70  & 18.70  & 11.50 & 9.40  & 10.60 & 12.00 & & 30.50  & 18.50 & 12.10 & 9.80  & 10.60 & 11.90 & & 35.80  & 24.50  & 16.20 & 10.40 & 8.50  & 10.20  \\
    & MCS-D     & 18.80  & 9.50   & 5.60  & 3.50  & 3.30  & 3.60  & & 18.40  & 8.80  & 5.00  & 3.40  & 3.30  & 3.50  & & 24.60  & 14.80  & 9.00  & 5.60  & 4.70  & 6.60   \\
    & LMD ($\lambda_{b}=15$) & 17.30  & 6.70  & 2.90  & \textbf{1.50 } & \textbf{1.90 } & \textbf{2.00 } & & 15.80  & 6.40  & 2.30  & \textbf{1.50 } & \textbf{1.60 } & 2.30  & & 20.30  & 11.20  & 4.60  & 2.10  & 0.90  & \textbf{1.30 } \\
    & LMD\; ($\lambda_{b}=0$)\;  & \textbf{8.80 } & \textbf{5.00 } & 5.40  & 8.80  & 11.50  & 13.60  & & \textbf{9.90 } & 6.40  & 6.30  & 9.40  & 12.50  & 15.50  & & \textbf{12.90 } & \textbf{6.00 } & 3.10  & 3.30  & 6.50  & 12.20  \\
    \bottomrule[1.3pt]
    \end{tabular}}
    \label{tab:detectors_perf_whitebox}
\end{table*}

\begin{table*}[t]
    \centering
    \caption{Detection EER of the detectors against three attackers in the black-box attack scenario on the two ASVs.}
    \scalebox{0.85}{\begin{tabular}{p{60pt}<{\centering}c
                                    *{6}{p{12pt}<{\centering}}   p{2pt}<{\centering}
                                    *{6}{p{12pt}<{\centering}}   p{2pt}<{\centering}
                                    *{6}{p{12pt}<{\centering}}}
    \toprule[1.3pt]
    EER (\%) $\searrow$ & Attacker $\rightarrow$ & \multicolumn{6}{c}{BIM ($L_{\infty}$, $\alpha=1$)} &  & \multicolumn{6}{c}{PGD ($L_2$, $\alpha=300$)} &  & \multicolumn{6}{c}{CW (RMS, $N=100$)} \\
    \specialrule{0em}{3pt}{3pt}
    Victim Model $\downarrow$ & $N / \kappa\; \rightarrow$ & 5     & 10    & 20    & 50    & 100   & 200 &   & 5     & 10    & 20    & 50    & 100   & 200 &  & 0     & 0.1   & 0.2   & 0.3   & 0.4   & 0.5 \\
    \midrule[0.8pt]

    \multirow{7}{*}{\makecell{ECAPA\_TDNN + \\ AAM-Softmax}}
    & Vocoder   & 49.60  & 50.60  & 47.30  & 43.80  & 41.20  & 41.10  & & 50.60  & 49.20  & 49.00  & 44.40  & 41.90  & 42.40  & & 49.80  & 50.50  & 49.70  & 49.60  & 50.90  & 48.20  \\
    & GL-mel    & 54.70  & 55.10  & 56.70  & 54.10  & 52.50  & 52.20  & & 54.50  & 56.30  & 57.00  & 55.20  & 54.70  & 53.20  & & 53.30  & 54.30  & 54.60  & 54.40  & 53.40  & 53.00  \\
    & GL-lin    & 53.30  & 53.50  & 51.00  & 45.50  & 45.00  & 43.10  & & 54.30  & 54.30  & 51.80  & 48.90  & 46.90  & 43.80  & & 53.40  & 54.30  & 55.70  & 55.30  & 53.50  & 51.60  \\
    & MCS-H     & 53.80  & 52.50  & 51.80  & 52.80  & 52.70  & 51.10  & & 53.00  & 53.10  & 53.00  & 52.90  & 52.60  & 51.20  & & 52.90  & 53.50  & 53.20  & 54.10  & 54.20  & 52.30  \\
    & MCS-D     & 52.10  & 50.40  & 45.80  & 39.60  & 37.80  & 36.10  & & 52.80  & 51.20  & 46.80  & 40.80  & 38.50  & \textbf{37.20 } & & 52.30  & 52.50  & 52.30  & 52.40  & 51.00  & 50.10  \\
    & LMD ($\lambda_{b}=15$) & 53.50  & 51.40  & 47.60  & 42.00  & 39.00  & 39.80  & & 53.30  & 51.90  & 47.80  & 42.80  & 40.50  & 39.70  & & 54.10  & 54.00  & 54.30  & 53.50  & 51.70  & 51.00  \\
    & LMD\; ($\lambda_{b}=0$)\; & \textbf{46.60 } & \textbf{43.70 } & \textbf{40.00 } & \textbf{36.30 } & \textbf{36.50 } & \textbf{36.00 } & & \textbf{45.80 } & \textbf{44.00 } & \textbf{41.20 } & \textbf{40.00 } & \textbf{39.70 } & 38.30  & & \textbf{47.60 } & \textbf{46.10 } & \textbf{44.40 } & \textbf{43.40 } & \textbf{43.50 } & \textbf{44.60 } \\
    \midrule[0.8pt]

    \multirow{7}{*}{\makecell{Fast-ResNet34 + \\ Angular Prototypical}}
    & Vocoder   & 51.60  & 48.60  & 46.40  & 44.80  & 44.00  & 44.30  & & 50.80  & 49.60  & 47.50  & 45.70  & 45.60  & 44.20  & & 50.40  & 50.20  & 51.00  & 49.20  & 49.90  & 50.00  \\
    & GL-mel    & 52.50  & 51.70  & 50.90  & 46.70  & 47.60  & 47.80  & & 51.90  & 51.40  & 50.50  & 48.30  & 48.10  & 48.50  & & 52.20  & 52.60  & 53.10  & 52.50  & 52.30  & 51.50  \\
    & GL-lin    & 49.00  & 47.10  & 42.60  & 38.20  & 39.50  & 40.00  & & 50.30  & 48.70  & 45.40  & 41.90  & 40.50  & 40.80  & & 50.00  & 50.30  & 48.90  & 48.10  & 48.50  & 47.30  \\
    & MCS-H     & 50.60  & 50.80  & 50.50  & 49.90  & 51.20  & 51.90  & & 51.40  & 50.80  & 51.20  & 51.40  & 51.90  & 52.60  & & 51.80  & 52.00  & 51.80  & 51.50  & 52.00  & 51.50  \\
    & MCS-D     & 50.50  & 48.70  & 42.90  & 38.00  & \textbf{36.50 } & \textbf{37.90 } & & 50.20  & 49.10  & 44.10  & 39.80  & \textbf{37.20 } & \textbf{38.70 } & & 50.30  & 49.50  & 48.80  & 47.80  & 46.60  & 45.10  \\
    & LMD ($\lambda_{b}=15$) & 50.80  & 48.90  & 43.70  & 39.60  & 39.60  & 41.10  & & 51.50  & 49.40  & 46.70  & 40.00  & 40.40  & 40.90  & & 53.50  & 54.30  & 52.30  & 50.60  & 48.60  & 49.00  \\
    & LMD\; ($\lambda_{b}=0$)\; & \textbf{43.80 } & \textbf{40.60 } & \textbf{38.30 } & \textbf{36.90 } & 38.30  & 38.60  & & \textbf{46.10 } & \textbf{43.30 } & \textbf{40.70 } & \textbf{39.70 } & 38.90  & 40.60  & & \textbf{47.20 } & \textbf{47.10 } & \textbf{45.30 } & \textbf{44.10 } & \textbf{43.70 } & \textbf{42.80 } \\
    \bottomrule[1.3pt]
    \end{tabular}}
    \label{tab:detectors_perf_blackbox}
\end{table*}

\begin{figure*}[t]
    \centering
    \subfigure[BIM ($L_{\infty}$)]{
        \includegraphics[width=0.33\linewidth]{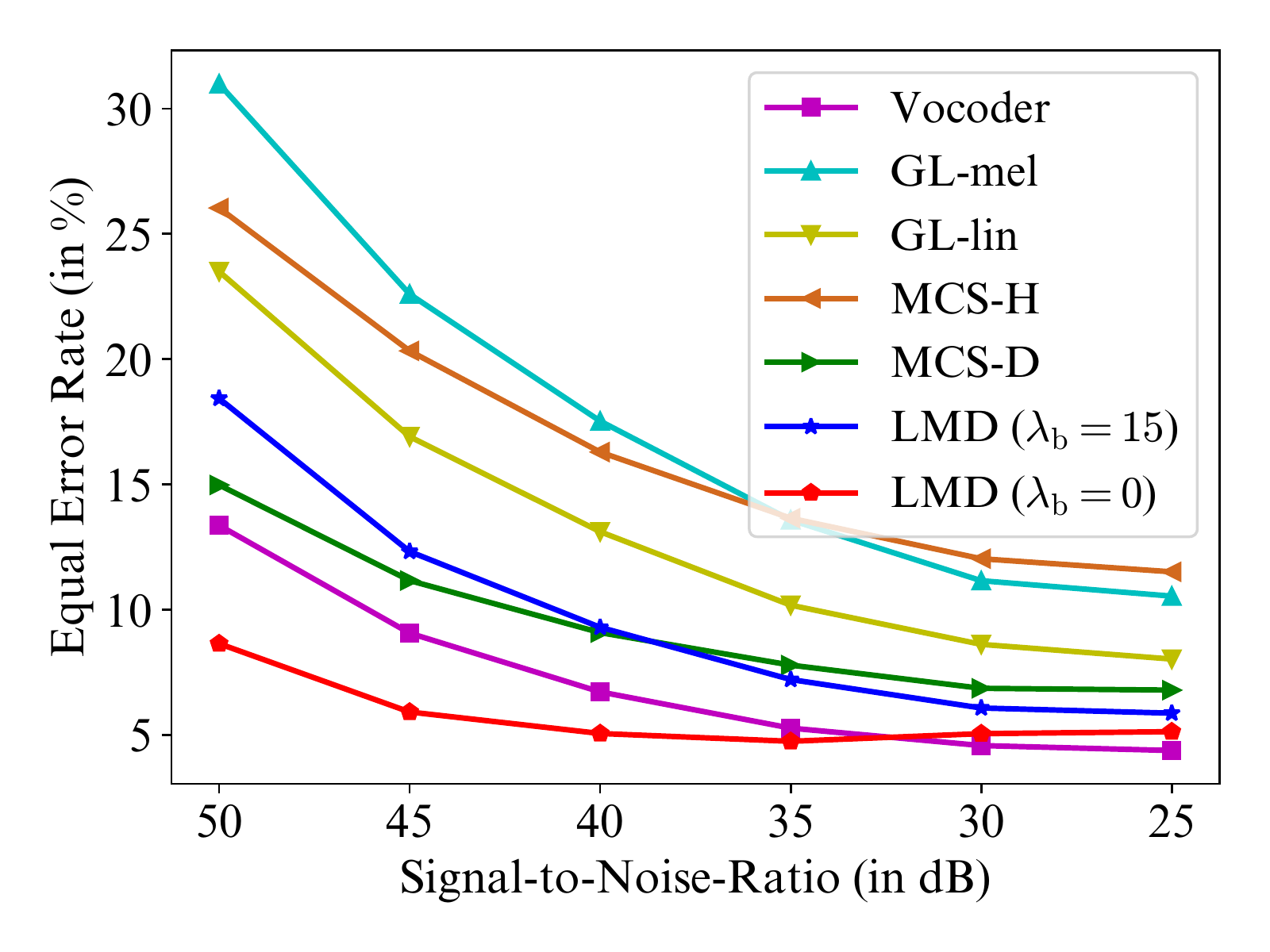}
    }
    \hspace{-16pt}
    \subfigure[PGD ($L_2$)]{
        \includegraphics[width=0.33\linewidth]{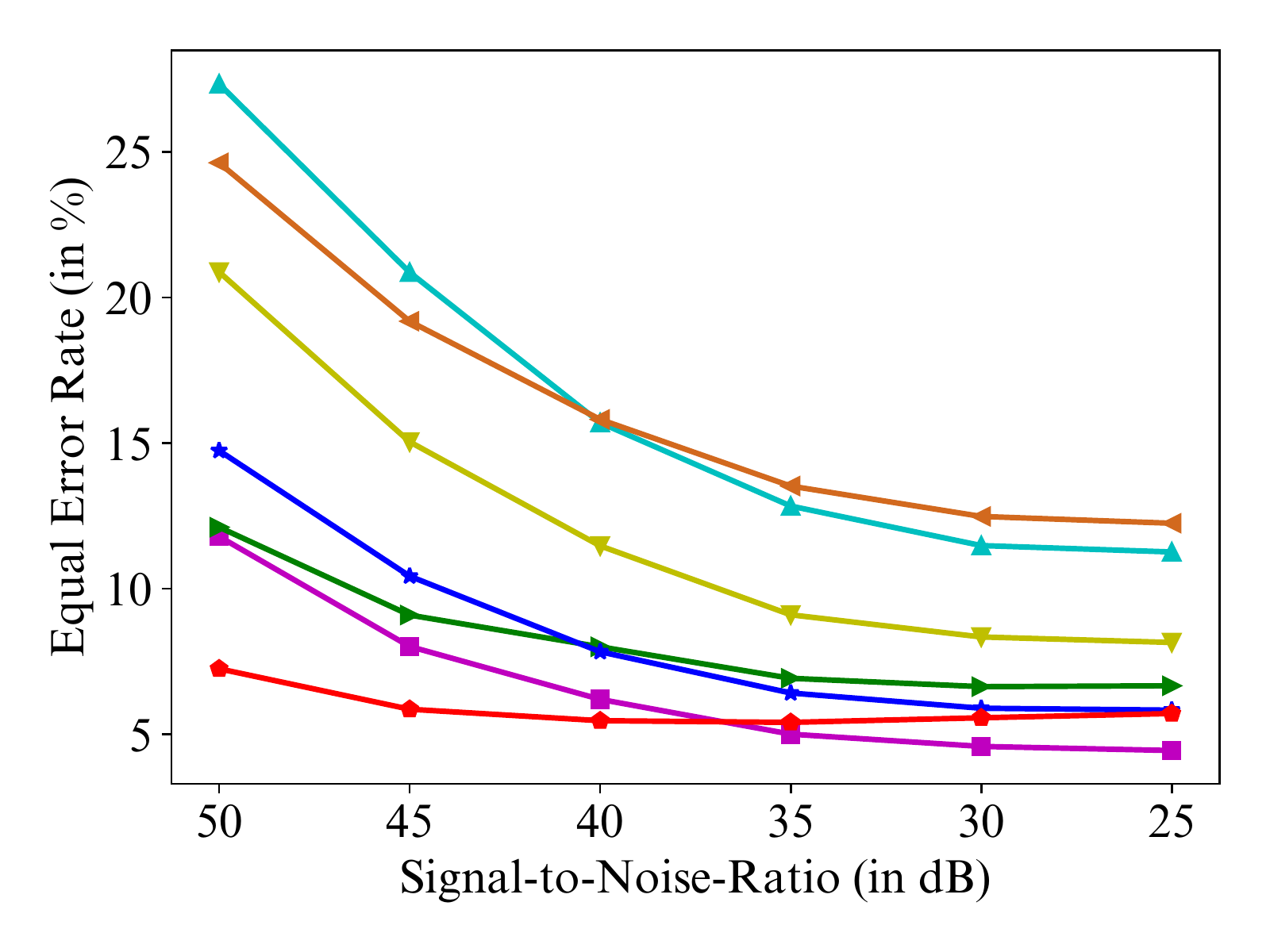}
    }
    \hspace{-16pt}
    \subfigure[CW (RMS)]{
        \includegraphics[width=0.33\linewidth]{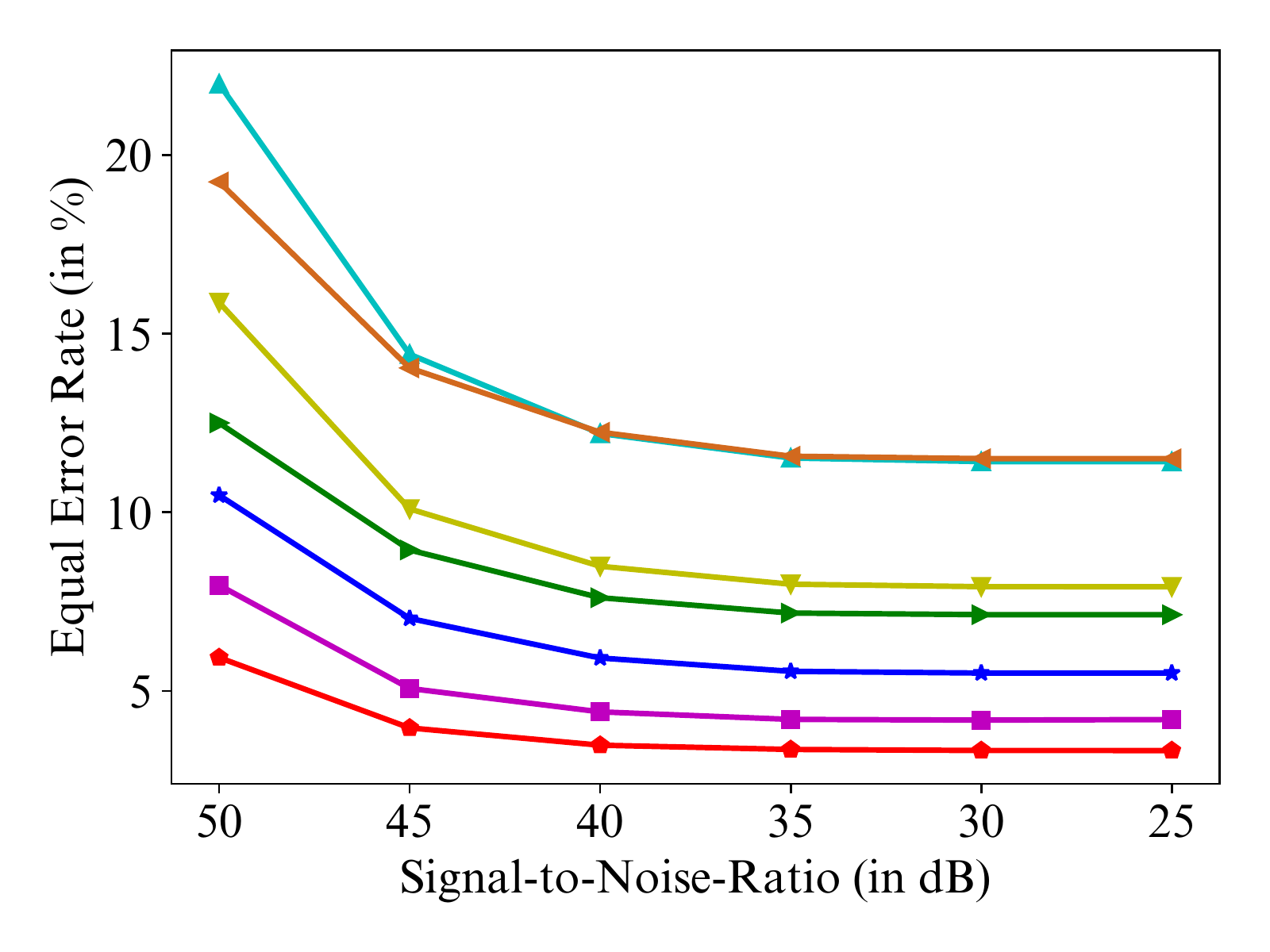}
    }
    \caption{Detection EER of the detectors along with the SNR budget. The performance of detectors was evaluated on ECAPA\_TDNN and the three adversaril mixture sets ($\mathcal{A}_{\text{\tiny BIM}}$, $\mathcal{A}_{\text{\tiny PGD}}$ and $\mathcal{A}_{\text{\tiny CW}}$) in the white-box attack scenario.}
    \label{fig:snrbudget_eer}
\end{figure*}

\section{Results and Discussions} \label{sec:exp_results}
In this section, we first present the performance of the attackers in Section~\ref{subsec:per_attack} so as to show their great threat to the ASV systems, then present the performance of the detectors against different attackers in Section~\ref{ssec:per_detector} so as to show how much the threat is mitigated. Finally, we present several additional experiments in Section~{\ref{ssec:abla_exps}}.

\subsection{Performance of The Attackers}\label{subsec:per_attack}

\figurename{\ref{fig:attackers_aenumber}} shows the number of adversarial examples at different ranges of SNR. The SNR of adversarial examples generated by the CW attacker is higher than that of the BIM and PGD attackers. Note that the SNRs are calculated on the adversarial mixture sets, i.e. $\mathcal{A}_{\text{\tiny BIM}}$, $\mathcal{A}_{\text{\tiny PGD}}$ and $\mathcal{A}_{\text{\tiny CW}}$.

\figurename{\ref{fig:attackers_perf}} illustrates the performance of the three attackers. ECAPA\_TDNN achieves an EER and minDCF of $1.25\%$ and $0.1372$ on the test list {\Courier VoxCeleb1-test}. Similarly, Fast-ResNet34 achieves $1.97\%$ and $0.2553$ respectively. The above results indicate that the two ASV models are SOTA. In the case of the white-box attacks, the BIM attacker and CW attacker achieves an ASR of 97\% at a SNR of 35dB and 42dB, respectively. The PGD attacker achieves similar performance with BIM. All of the three attackers leads to an minDCF of $0.99+$ even at a SNR of 45dB. In the case of the transfer-based black-box attack, the attackers generally deliver better performance on the Fast-ResNet34 ASV than on the ECAPA\_TDNN ASV. The BIM, PGD and CW attacker achieve their maximum ASR of 23\%, 20\% and 7\% on Fast-ResNet34, respectively. These results show that the attackers highly threaten the SOTA ASV models.

\begin{table}[t]
    \centering
    \caption{DSR of the detectors along with $\text{FAR}_{\mathrm{given}}$. The performance of detectors was evaluated on the adversarial mixture set of $\mathcal{A}_{\text{\tiny BIM}}$. The best results are in \textbf{bold}, while the runner-up results are \underline{underlined}.}
    \scalebox{0.85}{\begin{tabular}{cccccc}
    \toprule
    DSR (\%) & $\text{FAR}_{\text{given}} (\%)$  & 5.0   & 1.0   & 0.5   & 0.1  \\
    \midrule
    \multicolumn{1}{c}{\multirow{7}[0]{*}{\makecell{ECAPA\_TDNN + \\ AAM-Softmax}}}
        & Vocoder   & \textbf{95.98 } & \textbf{93.15 } & \textbf{92.02 } & \textbf{88.80 } \\
        & GL-mel    & 86.98  & 80.82  & 78.55  & 60.72  \\
        & GL-lin    & 90.47  & 83.35  & 80.08  & 69.22  \\
        & MCS-H     & 82.37  & 73.30  & 70.58  & 52.28  \\
        & MCS-D     & 91.73  & 78.97  & 71.50  & 66.15  \\
        & LMD ($\lambda_{b}=15$) & 93.92  & \underline{91.07}  & \underline{89.97}  & \underline{88.37}  \\
        & LMD \;($\lambda_{b}=0$)\; & \underline{94.85}  & 90.72  & 89.20  & 82.35  \\
    \midrule
    \multicolumn{1}{c}{\multirow{7}[1]{*}{\makecell{Fast-ResNet34 + \\ Angular Prototypical}}}
        & Vocoder   & \textbf{95.25 } & \textbf{91.05 } & \textbf{89.05 } & \textbf{85.67 } \\
        & GL-mel    & 88.72  & 82.73  & 80.10  & 69.63  \\
        & GL-lin    & 92.58  & 86.37  & 83.37  & 78.20  \\
        & MCS-H     & 74.98  & 66.82  & 62.90  & 51.53  \\
        & MCS-D     & 88.30  & 74.67  & 71.45  & 57.02  \\
        & LMD ($\lambda_{b}=15$) & \underline{93.38}  & \underline{89.50}  & \underline{88.67}  & \underline{84.07}  \\
        & LMD \;($\lambda_{b}=0$)\; & 89.28  & 82.03  & 79.68  & 72.17  \\
    \bottomrule
    \end{tabular}}
    \label{tab:dsr_givenFAR}
\end{table}

\begin{figure}[t]
    \centering
    \subfigure[ECAPA\_TDNN + AAM-Softmax]{
        \includegraphics[width=0.8\linewidth]{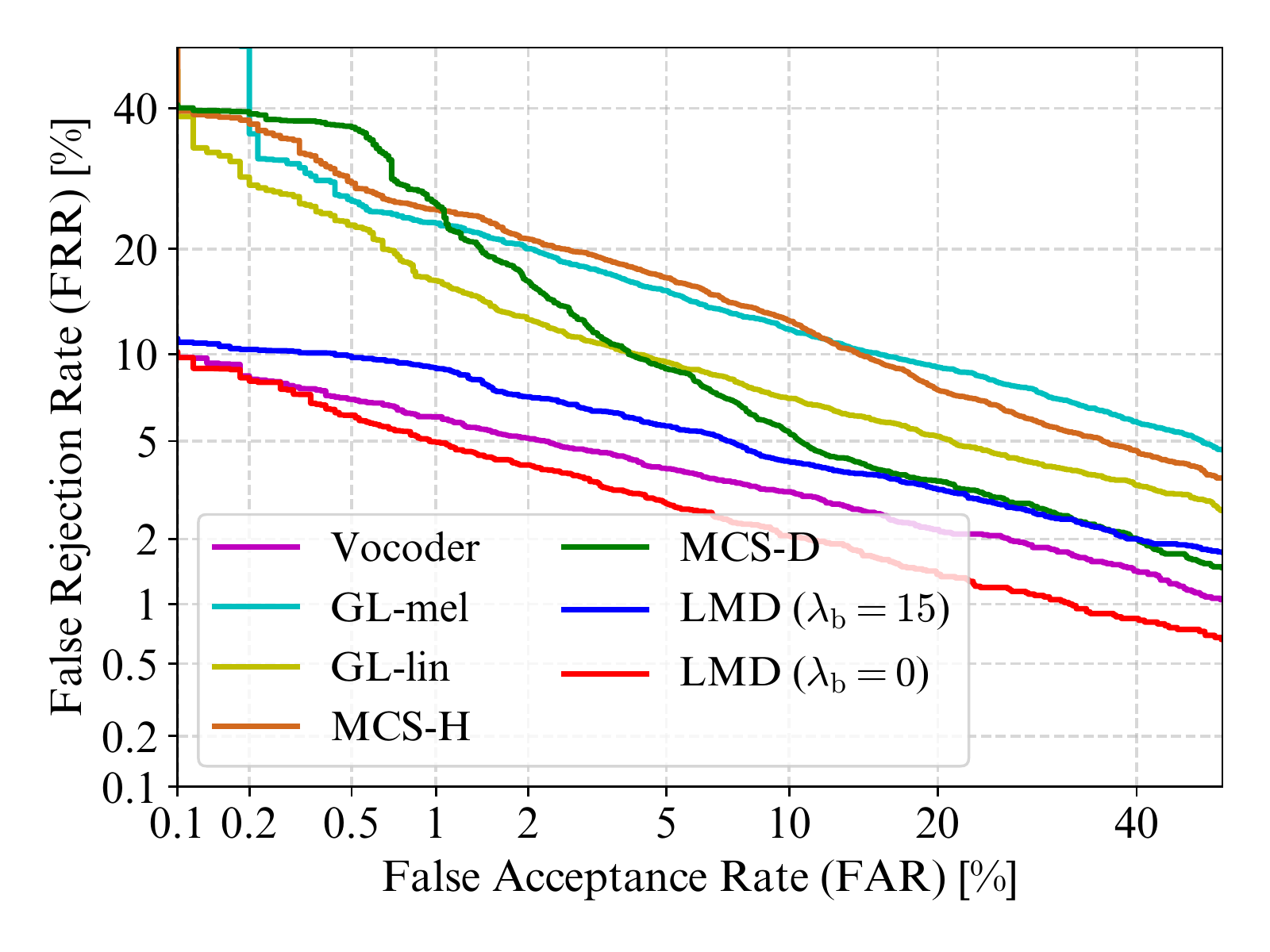}
    }
    \subfigure[Fast-ResNet34 + Angular Prototypical]{
        \includegraphics[width=0.8\linewidth]{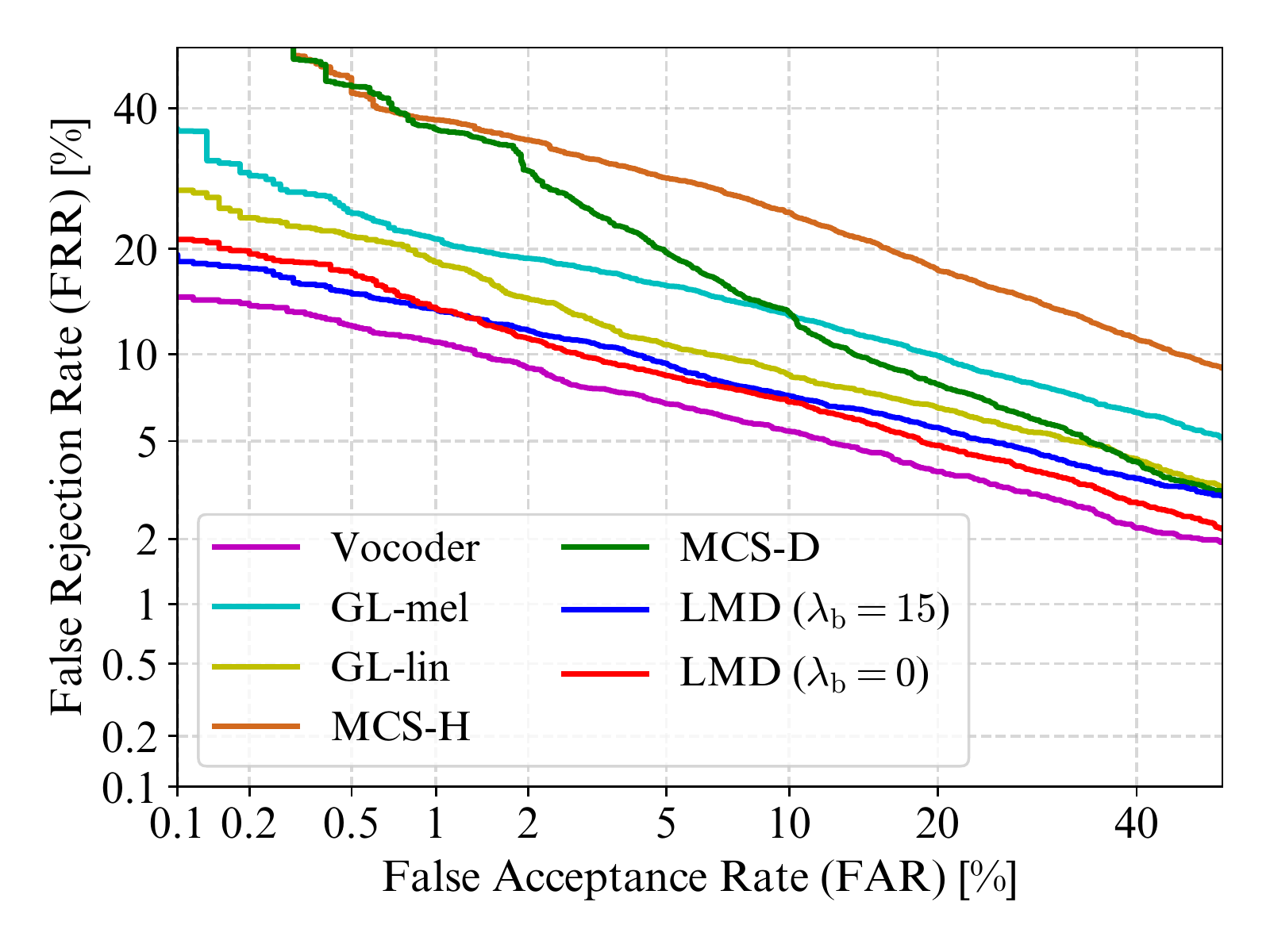}
    }
    \caption{DET curves of the detectors on the adversarial mixture set of $\mathcal{A}_{\text{\tiny CW}}$.}
    \label{fig:detectors_detcurve}
\end{figure}

\begin{figure}[t]
    \centering
    \subfigure[ECAPA\_TDNN + AAM-Softmax]{
        \includegraphics[width=0.8\linewidth]{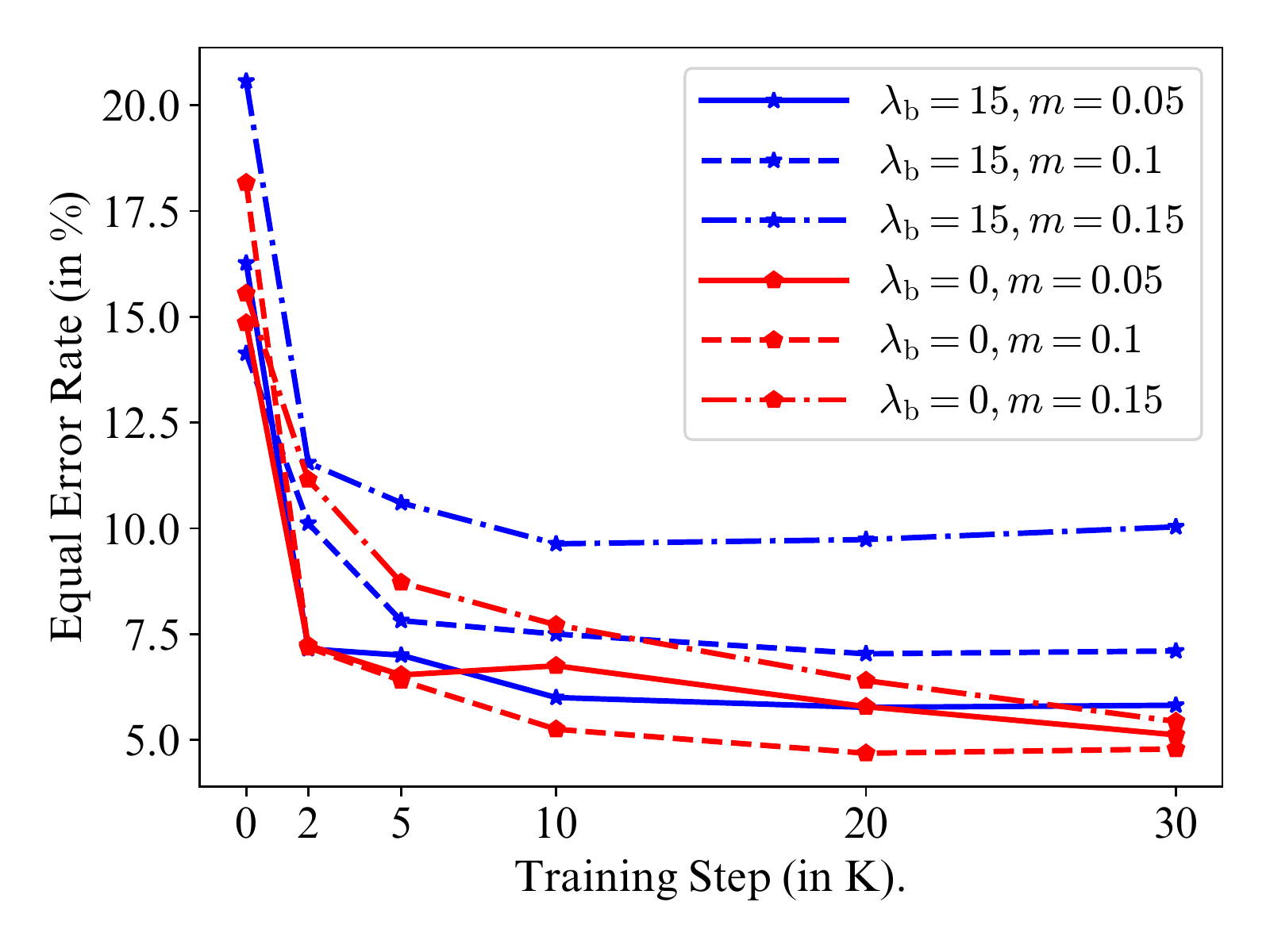}
    }
    \subfigure[Fast-ResNet34 + Angular Prototypical]{
        \includegraphics[width=0.8\linewidth]{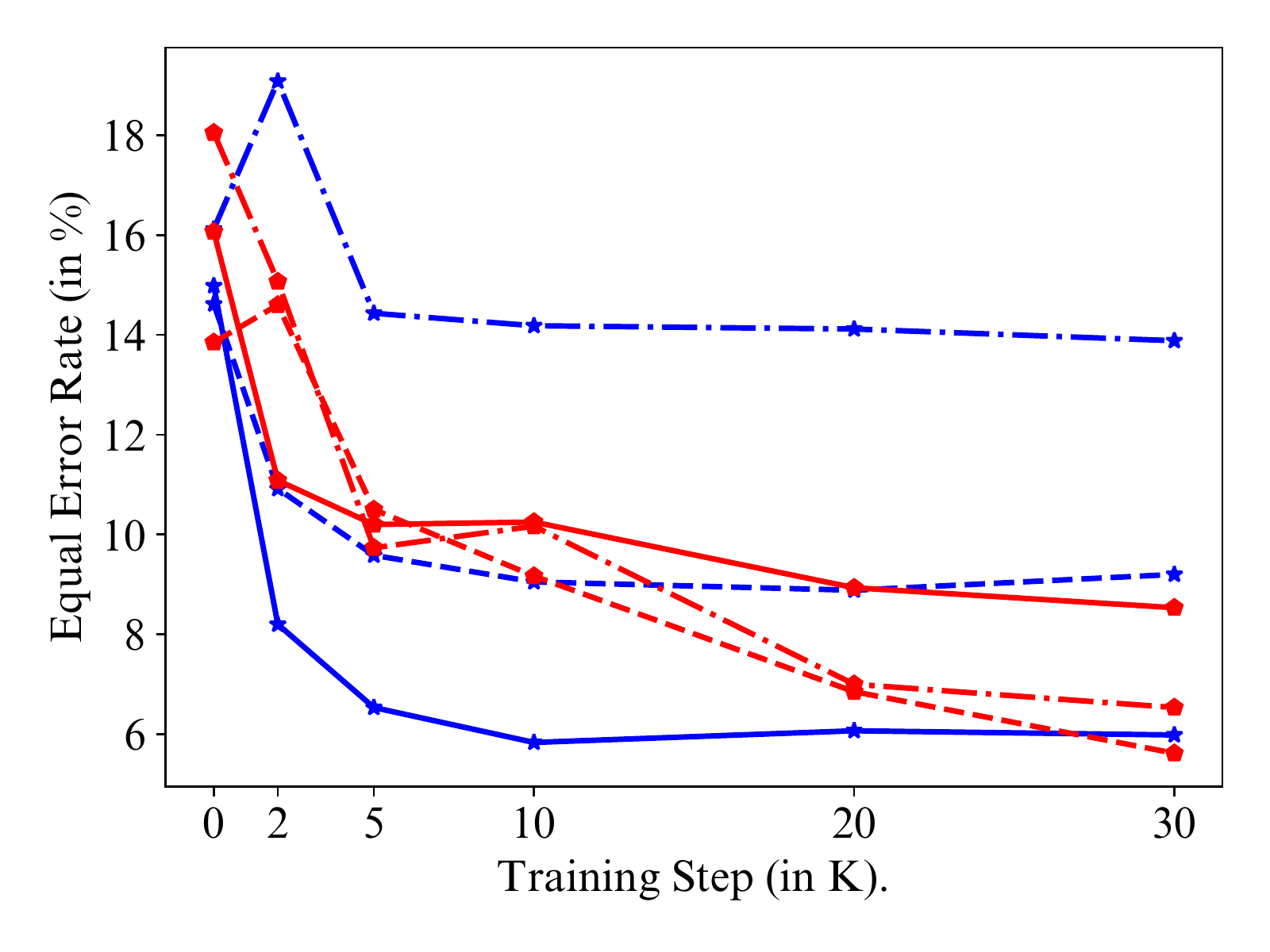}
    }
    \caption{Connection between the detection EER and training steps of our proposed LMD with different score margins, on the adversarial mixture set of $\mathcal{A}_{\text{\tiny BIM}}$.}
    \label{fig:training_step_eer}
\end{figure}

\subsection{Performance of The Detectors }\label{ssec:per_detector}
The performance of our proposed detectors is shown below, where $\lambda_{\text{b}}=15$ indicates the LMD-AIBM method, and $\lambda_{\text{b}}=0$ indicates the LMD-IRM method. The difference between the two methods lies in the type of their masking matrices.

Tables {\ref{tab:detectors_perf_whitebox}} and {\ref{tab:detectors_perf_blackbox}} comprehensively show the EER performance of the detectors in the white-box and black-box scenarios, respectively. Note that the EER is calculated in a noisy situation by evaluating $\mathbf{E}\bigl(\mathcal{A}_{i}, \mathcal{G}_{i}\bigr)$ for the three attackers, where $i=1,2,\cdots,6$ represent the six different parameter settings. The victim ASV and the defended ASV are always consistent. Several conclusions can be drawn: (i) From the perspective of the white-box attack scenario, our proposed LMD method outperforms the baseline methods in the most detection conditions. For example, LMD-AIBM achieves a detection EER of 0.8\% and 1.5\% on ECAPA\_TDNN and Fast-ResNet34, respectively, when encountering the BIM attacker with $N=50$, which is 38\% and 11\% higher than Vocoder. (ii) MCS-H possesses the worst detection performance due to its coarse mask matrix, while MCS-D achieves comparable performance to GL-mel and GL-lin by finely designing the mask matrix, which shows the effectiveness of our mask-based idea, despite the mask matrices of MCS-H and MCS-D are both manually crafted. (iii) Further, we desire to leverage the neural network to learn an AIBM matrix or an IRM matrix for detection.
LMD-AIBM performs better than LMD-IRM when the perturbation intensity is high, while LMD-IRM performs better than LMD-AIBM when the perturbation intensity is low. (iv) From the perspective of the black-box attack scenario, our proposed LMD method achieves the optimal performance in almost all detection conditions. The results on ECAPA\_TDNN and Fast-ResNet34 are basically the same, obtaining an EER of 37\% when encountering the BIM attacker and the PGD attacker, and an EER of 44\% when encountering the CW attacker. There is still great development potential to sperate adversarial examples in the black-box scenario for the detection-based passive defense approaches.
In addition, we believe that the main reason for the low detection performance of black-box attacks is that the large number of failed adversarial examples pulls down the adversarial score variation, thus leading to higher detection EER. Therefore, we conducted an ablation experiment in Section \ref{sssec:realAE-ablation}.

\figurename{\ref{fig:snrbudget_eer}} shows the impact of the SNR budget on the detector performance. From the figure, we draw the following conclusions: (i) the performance of all detectors gradually drops as the SNR budget decreases. In the range of SNR budget of 50dB to 40dB, our LMD-IRM maintains an EER of 3\% to 9\% and outperforms the comparison methods. (ii) In the range of SNR budget of 35dB to 25dB, our LMD methods achieve comparable performance with Vocoder with an EER fluctuating around 5\%. (iii) Our proposed LMD-IRM outperforms all baseline detectors at a SNR budget higher than 37 dB.

\tablename{\ref{tab:dsr_givenFAR}} shows the variation of the detector accuracy with the $\text{FAR}_{\mathrm{given}}$. From the figure, it can be concluded that, as the $\text{FAR}_{\mathrm{given}}$ decreases from 5\% to 0.1\%, the detection threshold will increase meanwhile, and more adversarial examples will be missed, so the accuracy of all detectors drops. Moreover, Vocoder reaches the top accuracy while its DSR drops from 96\% to 86\%, on the contrary, our proposed LMD-AIBM achieves the runner-up accuracy while its DSR drops from 93\% to 84\%.

In \figurename{\ref{fig:detectors_detcurve}}, the detection error tradeoff (DET) curve is used to evaluate the detector performance more delicately than \tablename{\ref{tab:dsr_givenFAR}}. Experimental results on the ECAPA\_TDNN ASV system indicate that our LMD-IRM detector is always ahead of Vocoder, and both of them obtain an EER of less than 5\%. Experimental results on the Fast-ResNet34 ASV system show that Vocoder always leads our LMD detectors with an EER of less than 10\%.

\begin{figure}[t]
    \centering
    \includegraphics[width=0.9\linewidth]{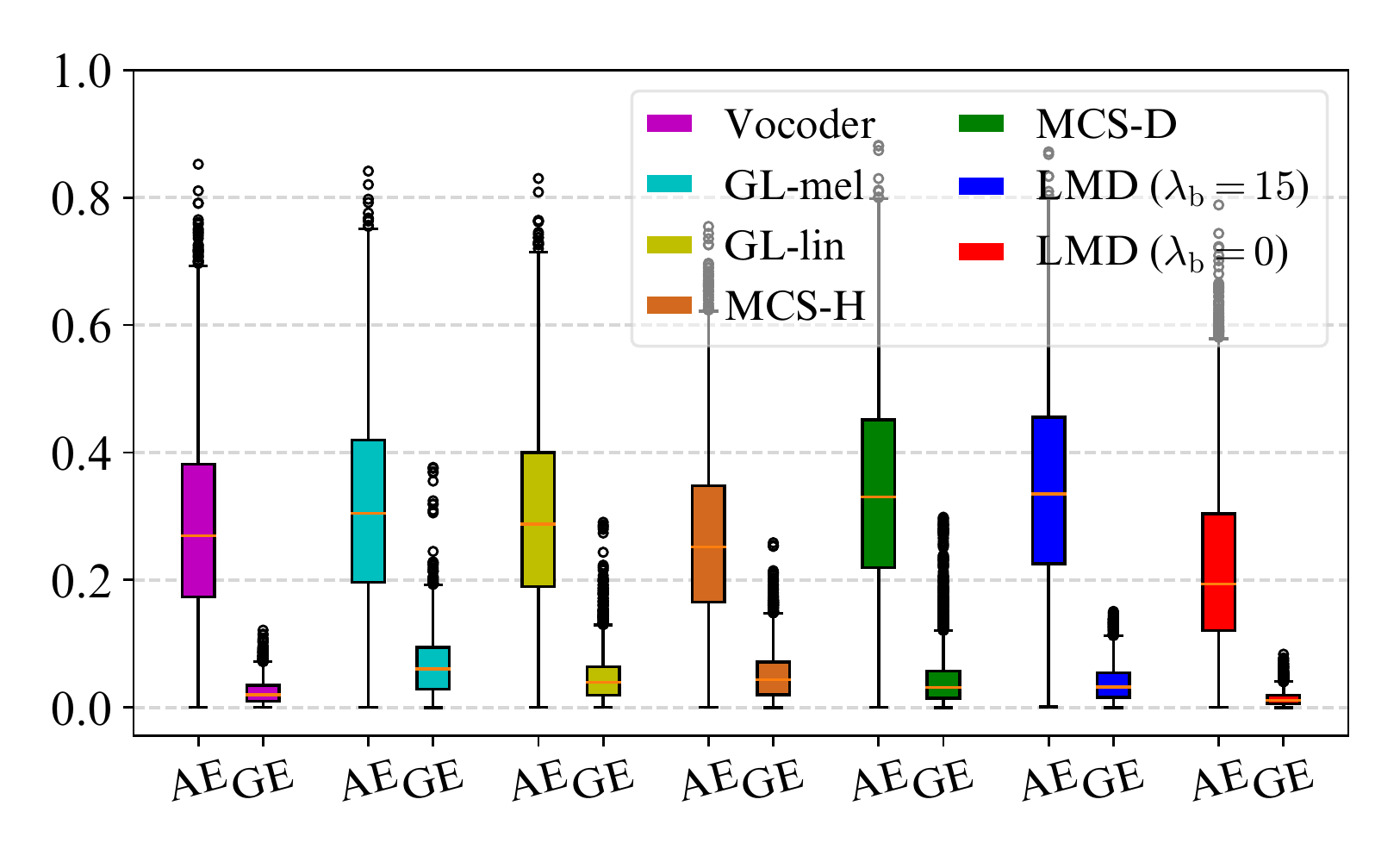}
    \caption{Boxplot of the score variations of the adversarial mixture set $\mathcal{A}_{\text{\tiny CW}}$ and genuine mixture set $\mathcal{G}_{\text{\tiny CW}}$ for the seven detectors with the ECAPA\_TDNN ASV as the victim. AE and GE represent adversarial examples and genuine examples, respectively.}
    \label{fig:detectors_boxplot}
\end{figure}

\begin{figure}[t]
    \centering
    \subfigure{
        \includegraphics[width=0.95\linewidth]{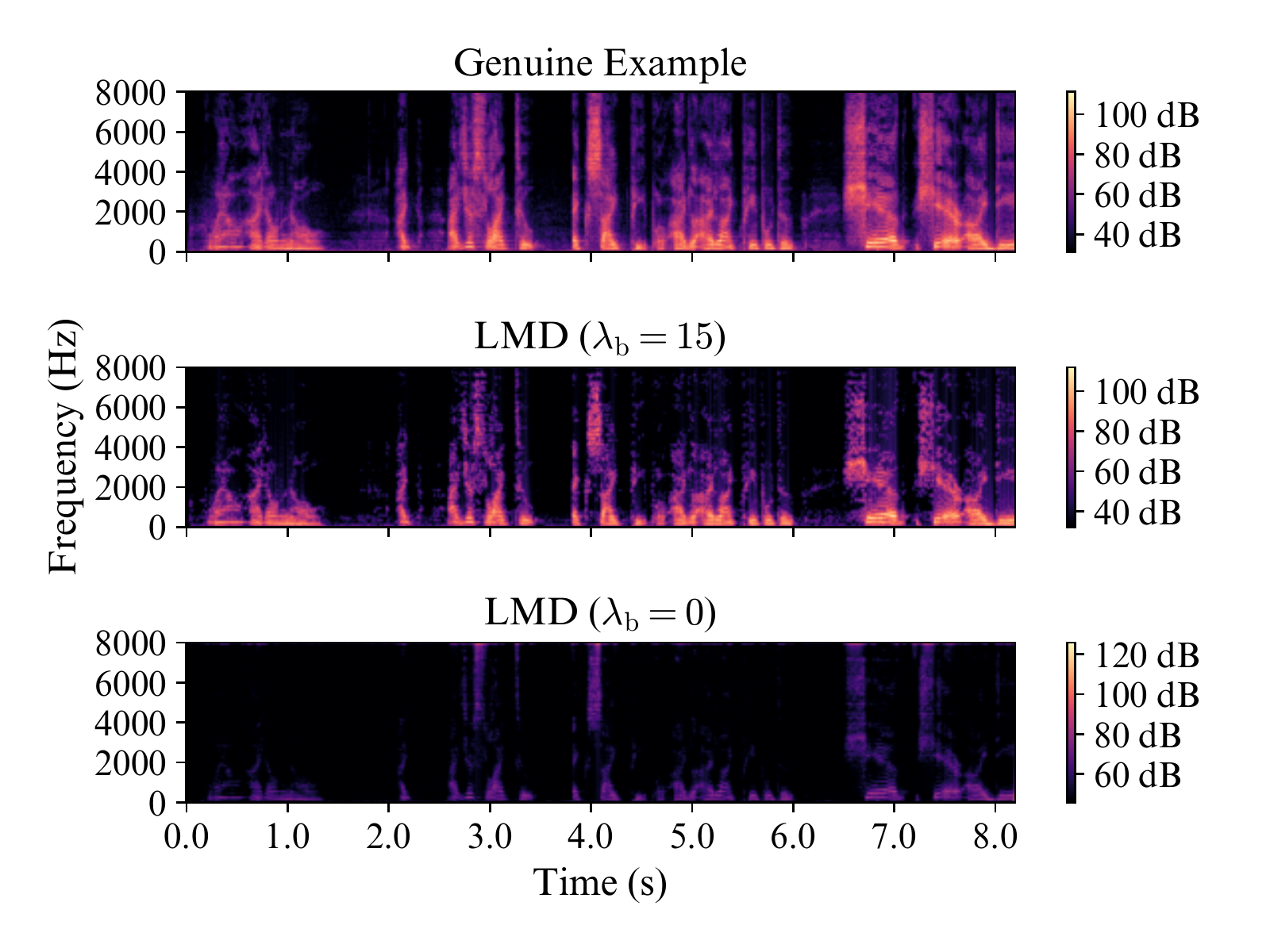}
    }
    \subfigure{
        \includegraphics[width=0.95\linewidth]{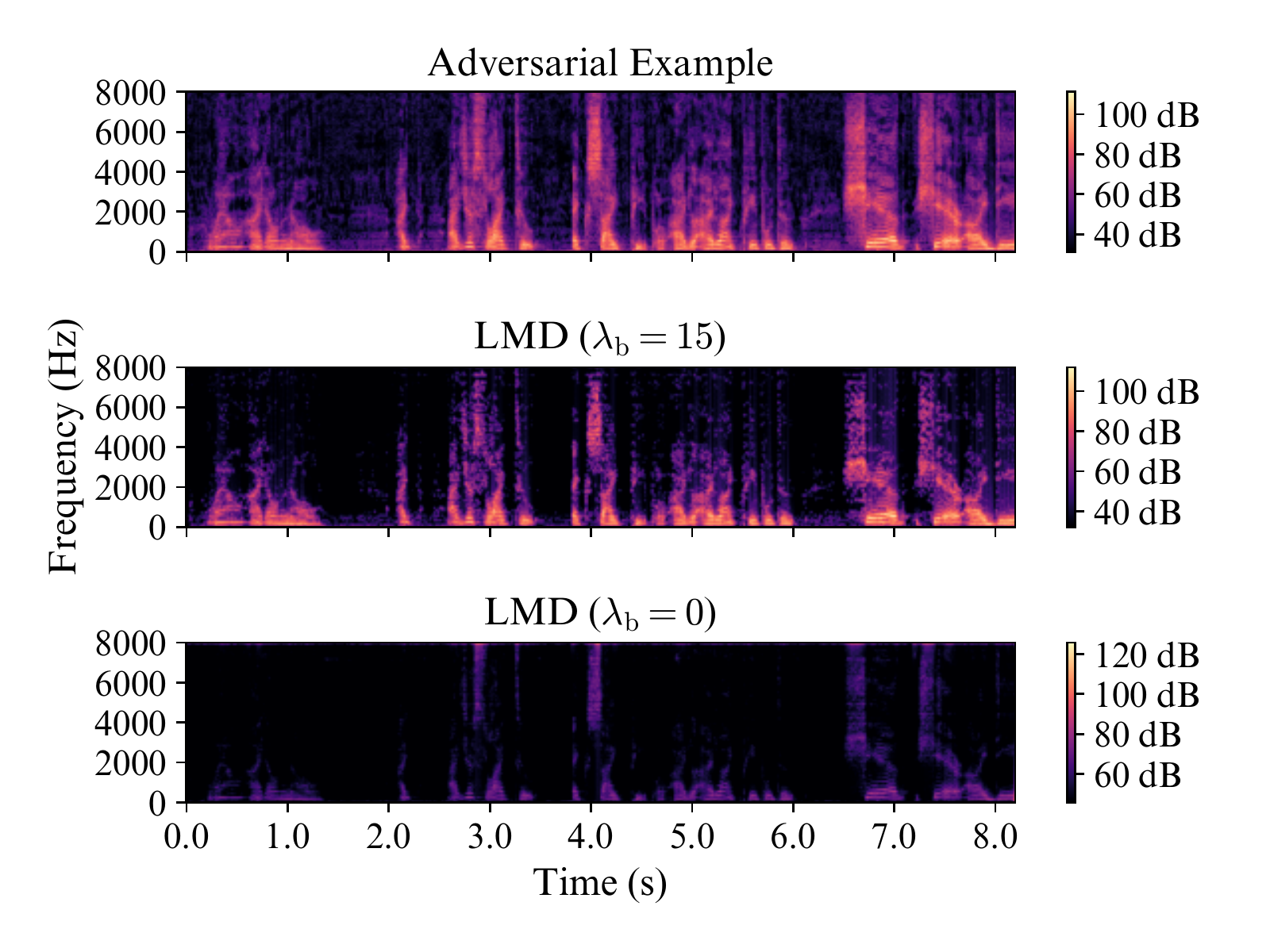}
    }
    \caption{Spectrograms of the original audio examples and their corresponding transformed audio examples obtained by our LMD. The genuine example is from {\Courier id10284/7yx9A0yzLYk/00010.wav} of VoxCeleb1. The hypothesis enrollment utterance of the adversarial example is from {\Courier id10305/gbTZ7k9e/Z0\_00001.wav} of VoxCeleb1. The adversarial example was generated by the BIM attacker with $N=50$.}
    \label{fig:clean_adver_specs}
\end{figure}

\subsection{Ablation Studies}\label{ssec:abla_exps}
\subsubsection{Effects of The Hyperparameter $m$ On Performance}
The hyperparameter $m$ in \eqref{equ:loss_term_variation}, i.e. the score margin, controls the amount of speaker information to be masked out. To study its effect on the performance of LMD, we trained the LMD-AIBM and LMD-IRM detectors with $m$ set to 0.05, 0.1 and 0.15, respectively, and evaluated them with $\mathcal{A}_{\text{\tiny BIM}}$ on the two ASVs. We draw several conclusions from the results in \figurename{\ref{fig:training_step_eer}} as follows. (i) In the initial naive state where a random mask matrix is generated, LMD obtains an EER of about 16\%, which proves the effectiveness of  our mask-based idea again. (ii) When the training of LMD proceeds, the detection EER gradually decreases and becomes smooth after 20K steps with an EER of 5\% to 8\%. (iii) For LMD-AIBM (blue lines) and LMD-IRM (red lines), the hyperparameter $m$ performs optimally on 0.05 and 0.1, respectively. However, we set $m$ to 0.05 in all experiments for the sake of controlling variables. (iv) LMD-AIBM is difficult to be trained successfully when $m$ is set large, especially when Fast-ResNet34 acts as the victim model, which could be mitigated by increasing $\lambda_{\text{b}}$.

\subsubsection{Interpretation of The Principles of LMD}
To explain why our proposed LMD method is effective, we present the boxplot of the score variations in \figurename{\ref{fig:detectors_boxplot}} for the analysis. Specifically, the boxplot depicts the distribution of the score variations of the detectors when confronted with the adversarial examples and genuine examples. From the figure, it can be seen that, our LMD ensures that the score variations for the genuine examples do not exceed $m$, and moreover, it makes the score variations for the adversarial examples as large as possible, which consequently gets the detection easier.

\figurename{\ref{fig:clean_adver_specs}} further visualizes the spectrograms of the original audio and transformed audio.
From the figure, it can be seen that LMD-AIBM masks the low-energy regions and samples sparsely, while LMD-IRM masks most of the low-energy regions, which are consistent with our goal of masking the most time-frequency bins that contain little speaker information. Therefore, they reach large score variations for the adversarial examples, and small score variations for the genuine examples.

\subsubsection{Purification Effects of LMD}

\begin{table}[t]
    \centering
    \caption{Accessional purification effects of our LMD.}
    \scalebox{0.85}{\begin{tabular}{
        p{60pt}<{\centering}c *{4}{p{20pt}<{\centering}}
    }
    \toprule
    \multirow{2}{*}{EER (\%) $\searrow$} & \multirow{2}{*}{Attacker $\rightarrow$}
    & \multirow{2}{*}{Clean} & BIM  & PGD & CW  \\
    & &  & ($L_{\infty}$) & ($L_2$)  & (RMS) \\
    \specialrule{0em}{2pt}{2pt}
    Victim Model $\downarrow$ & $N/\kappa \rightarrow$ & - & 200   & 200   & 0.5 \\
    \midrule
    \multicolumn{1}{c}{\multirow{4}[2]{*}{\makecell{ECAPA\_TDNN + \\ AAM-Softmax}}}
        & No-Defense & 1.25  & 100.00  & 99.80  & 98.40  \\
        & Vocoder & \textbf{1.20 } & 70.80  & 64.40  & 11.80  \\
        & LMD ($\lambda_{\text{b}}=15$) & 1.80  & \textbf{30.00 } & \textbf{25.00 } & \textbf{3.60 } \\
        & LMD ($\lambda_{\text{b}}=0$) & 1.40  & 97.60  & 97.20  & 42.00  \\
    \midrule
    \multicolumn{1}{c}{\multirow{4}[2]{*}{\makecell{Fast-ResNet34 + \\ Angular Prototypical}}}
        & No-Defense & 1.97  & 100.00  & 100.00  & 10.00  \\
        & Vocoder & \textbf{1.80 } & 58.40  & 52.80  & 21.20  \\
        & LMD ($\lambda_{\text{b}}=15$) & 2.80  & \textbf{20.00 } & \textbf{18.80 } & \textbf{8.20 } \\
        & LMD ($\lambda_{\text{b}}=0$) & 2.00  & 99.40  & 98.80  & 85.00  \\
    \bottomrule
    \end{tabular}}
    \label{tab:purification_effects}
\end{table}

The PWG-based Vocoder has also been utilized for the mitigation-based defense in \cite{joshi2021study}. Here we further explored the effectiveness of our LMD to purify the adversarial noise in Table \ref{tab:purification_effects}, where we used the pre-trained model provided by \textit{Wu et al.} \cite{wu2022adversarial} as a baseline, and employed EER of the victim ASV as the evaluation metric. From the table, we see that LMD-AIBM achieves much better purification effect than LMD-IRM, because the $L_1$ norm of the mask matrix measures the masking degree of LMD-AIBM more appropriately than LMD-IRM. Our LMD is designed to mask as many spectrogram bins as possible at the cost of little speaker information. Therefore, EER increases slightly on clean examples but decreases the most on adversarial examples. However, Vocoder behaves more like a speech enhancement module, where the input goes through a front-end noise reduction. Therefore, EER decreases on clean examples but the decrease in EER on adversarial examples is less apparent than LMD. Compared with Vocoder, the threat brought by adversarial examples are significantly mitigated by our LMD-AIBM.

\begin{table}[t]
    \centering
    \caption{Detector performance against adaptive attackers. The term ``trans (w/o)'' means that the input is not transformed by the detector, while the term ``trans (w/)'' is the opposite. ECAPA\_TDNN is employed as the victim ASV.}
    \scalebox{0.95}{\begin{tabular}{ccccc}
    \toprule
    \multicolumn{1}{c}{\multirow{2}{*}{Attacker $\downarrow$}}
    & \multicolumn{1}{c}{\multirow{2}{*}{Detector $\downarrow$}}
    & \multicolumn{2}{c}{ASR (\%)}
    & \multicolumn{1}{c}{\multirow{2}{*}{EER (\%)}} \\
    & & trans. (w/o) & trans. (w/) &  \\
    \midrule
    BIM                 & Vocoder & 69.20  & 83.70  & 12.50  \\
    ($L_{\infty},N=50$) & LMD-AIBM & 3.50  & 82.80  & \textbf{1.70 } \\
    \midrule
    PGD                 & Vocoder & 64.70  & 80.80  & 12.10  \\
    ($L_{2},N=50$) & LMD-AIBM & 2.80  & 73.90  & \textbf{2.10 } \\
    \midrule
    CW                  & Vocoder & 41.60  & 84.80  & 9.80  \\
    (RMS$,\kappa=0.3$)  & LMD-AIBM & 1.70  & 42.30  & \textbf{5.60 } \\
    \bottomrule
    \end{tabular}}
    \label{tab:adaptive-exps}
\end{table}

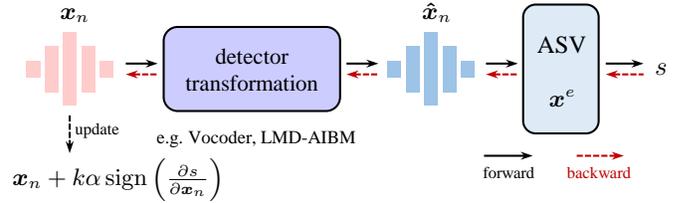
\begin{figure}[t]
    \centering
    \includegraphics[width=0.49\textwidth]{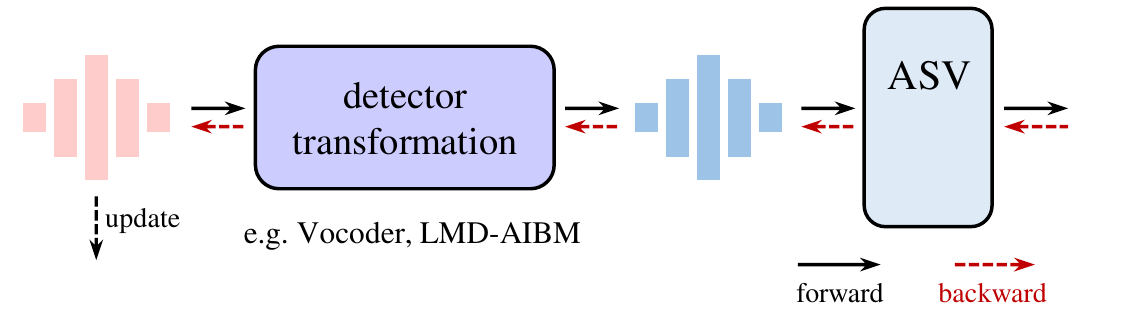}
    \put(-235,63){\small\bfseries\color{black}{$\bm{x}_n$}}
    \put(-98,63){\small\bfseries\color{black}{$\bm{\hat{x}}_n$}}
    \put(-50,30){\small\bfseries\color{black}{$\bm{x}^e$}}
    \put(-10,42){\normalsize\bfseries\color{black}{$s$}}
    \put(-253,0){\small\bfseries\color{black}{$\bm{x}_n + k\alpha \operatorname{sign}\left(\frac{\partial s}{\partial \bm{x}_n}\right)$}}
    \caption{Pipeline of the adaptive attacker for generating adversarial examples. The symbols $\bm{x}_n$ and $\bm{\hat{x}}_n$ denote the test utterance and the transformed utterance, respectively. The BIM attacker in Eq. \eqref{equ:bim_algorithm} is used here as an example.}
    \label{fig:adaptive-attacker}
\end{figure}

\begin{table}[t]
    \centering
    \caption{Detector performance under more realistic noises. The term ``LMD-AIBM (w/)'' indicates the training was augmented with the noises from MUSAN, and ``LMD-AIBM (w/o)'' indicates no data-augmentation. BIM ($L_{\infty}, N=200$) attacker is employed.}
    \scalebox{0.95}{\begin{tabular}{
            p{60pt}<{\centering}c *{4}{p{15pt}<{\centering}}
    }
    \toprule
    Victim Model $\downarrow$ & Noise Type $\rightarrow$ & gauss & noise & speech & music \\
    \midrule
    \multicolumn{1}{c}{\multirow{3}[2]{*}{\makecell{ECAPA\_TDNN + \\ AAM-Softmax}}}
        & Vocoder        & 2.00  & 1.90  & 1.80  & 1.80  \\
        & LMD-AIBM (w/o) & \textbf{1.20 } & 1.10  & \textbf{1.10 } & 1.40  \\
        & LMD-AIBM (w/)  & 1.60  & \textbf{1.00 } & \textbf{1.10 } & \textbf{1.00 } \\
    \midrule
    \multicolumn{1}{c}{\multirow{3}[2]{*}{\makecell{Fast-ResNet34 + \\ Angular Prototypical}}}
        & Vocoder        & \textbf{2.00 } & 2.00  & 2.30  & 2.40  \\
        & LMD-AIBM (w/o) & \textbf{2.00 } & 2.10  & 2.00  & 2.00  \\
        & LMD-AIBM (w/)  & 2.10  & \textbf{1.60 } & \textbf{1.90 } & \textbf{1.90 } \\
    \bottomrule
    \end{tabular}}
    \label{tab:noisetype-exps}
\end{table}

\subsubsection{Encounter With Adaptive Attackers}
\tablename{\ref{tab:adaptive-exps}} explores the performance of the detectors under the adaptive attack. The so-called adaptive attack means that the attacker can further access the detector parameters to generate an adversarial example. Specifically, as shown in \figurename{\ref{fig:adaptive-attacker}}, adversarial examples are updated by utilizing the gradient of the score w.r.t. the test utterance. From the \tablename{\ref{tab:adaptive-exps}}, two conclusions can be drawn: (i) the adaptive attacker cannot breach the system without a LMD-AIBM transformation, for example, the ASR drops from 82.8\% to 3.5\% under the BIM attacker. (ii) LMD-AIBM can also achieve a detection EER no higher than 5.6\% under the adaptive attack. Our analysis reveals that the attackers can only breach the original victim system, or the hybrid system with the LMD-AIBM transformation, i.e., they cannot breach both systems simultaneously. Eventually, we utilize the score variation of the two systems to detect the adversarial examples, and these adaptive adversarial examples will still yield a large score variation, so the detection performance remains robust.

\subsubsection{Data Augmentation for LMD}
\tablename{\ref{tab:noisetype-exps}} explores the detection performance of the LMD-AIBM with and without data-augmentation against a variety of noises. Specifically, we employ the MUSAN corpus \cite{snyder2015musan} for data-augmentation with a probability of 60\%. First, the noise sample is cropped or padded (in wrap mode) to the target length, and then it is scaled to a random SNR between [25, 40] before being superimposed to the target speech. The detection EER is employed as the evaluation metric, i.e., $\mathbf{E}\bigl(\mathcal{A}, \mathcal{G}\bigr)$, where $\mathcal{G}$ is constructed by adding noises to the original clean utterances, such as the Gaussian white-noise, or three types of noises from the MUSAN corpus. From the table, two conclusions can be drawn: (i) data-augmentation can further improve the detection performance of LMD-AIBM. (ii) The detection EER increases slightly for the type of noise that is unseen during the training, i.e., the Gaussian white-noise.

\begin{table}[t]
    \centering
    \caption{Performance of the detectors when applying the white-box adversarial attackers on the ASV victim systems (either calibrated, i.e. ``Cosine'', or un-calibrated, i.e. ``BCE''), where the attackers generate adversarial examples either from the calibrated ASV victim system or from the un-calibrated ASV victim system. The attacker BIM ($L_{\infty}, N=50$) and the victim ASV ECAPA\_TDNN are employed. The calibrated system obtains an ${\rm act.DCF}_{0.01}$ of 0.19 on the genuine examples.}
    \scalebox{0.85}{\begin{tabular}{
        c c cccc
    }
    \toprule
    \multicolumn{1}{c}{\multirow{2}{*}{\makecell{ASV victim \\ models} {$\downarrow$}}}
    &\multicolumn{1}{c}{\multirow{2}{*}{\makecell{White-box \\ Attackers} {$\downarrow$}}}
    & \multicolumn{2}{c}{Attacker Perf.} & \multicolumn{2}{c}{Detector Perf. by EER (\%)}  \\
    \specialrule{0em}{1pt}{1pt}

    & & ASR (\%)   & act.DCF & Vocoder & LMD-AIBM \\
    \midrule
    Cosine     & Cosine  & 95.80  & -     & 1.30  & 0.90  \\
    Cosine     & BCE   & 95.50  & -     & 1.50  & 1.10  \\
    \midrule
     BCE   & Cosine & 79.50  & 0.77  & 1.30  & 0.90  \\
     BCE   & BCE   & 78.90  & 0.78  & 1.50  & 1.10  \\
    \bottomrule
    \end{tabular}}
    \label{tab:calibration-exps}
\end{table}

\subsubsection{Effect of Calibration on Performance}
In the previous sections, we only studied the situation where the ASV victim systems are un-calibrated, i.e. they simply produce the similarity of two embeddings in terms of some measurement, like cosine similarity. However, the ASV systems are typically calibrated \cite{brummer2007fusion,brummer2013bosaris} in their real-world applications, i.e. they transform the un-calibrated similarity score of two embeddings to a target posterior probability, denoted as a calibrated score. A common calibration function is the binary-cross-entropy (BCE) loss \cite{villalba2020x}. In this section, we will further study the situation where the ASV victim systems are calibrated.

For each of the above two ASV victim systems, we can also have two kinds of white-box attackers: one kind generates adversarial examples from a victim system with the un-calibrated loss, such as the ``Cosine'' similarity $\mathbf{S}\left(\cdot\right)$ in Eq. \eqref{equ:bim_algorithm}, and the other kind generates adversarial examples from a victim system with the calibrated loss, such as BCE. Finally, we have four ``ASV-attacker'' pairs.

In this section, we present the performance of the detectors on the evaluation environments of the above four ``ASV-attacker'' pairs in \tablename{\ref{tab:calibration-exps}}. From the table, three conclusions can be drawn: (i) the calibration does not affect the detection EER, due to the fact that only positive scaling and offset are performed on the scores, whereas we obtain the variation of log-likelihood-ratio for detection. (ii) The ASR decreases after the calibration, because the threshold corresponding to EER and the threshold of the Bayesian decision operate on different points. (iii) The generation losses of ``Cosine'' and ``BCE'' produce the equivalent adversarial examples in terms of both principle and experimental results. They show little difference in terms of ASR, act.DCF and detection EER.

\begin{table}[t]
    \centering
    \caption{Detection EER of the detectors in the absence of those failed adversarial examples. The attacker BIM ($L_{\infty}$) and the victim ASV ECAPA\_TDNN are employed.}
    \scalebox{0.9}{\begin{tabular}{p{50pt}<{\centering} c *{6}{p{12pt}<{\centering}}}
    \toprule
    EER (\%) $\searrow $ & $N,\epsilon \rightarrow$
        & 5     & 10    & 20    & 50    & 100   & 200 \\
    \midrule
    \multirow{3}[2]{*}{White-box} & Vocoder & \textbf{2.74 } & 2.48 & \textbf{1.20 } & 1.15  & 1.72  & 1.91  \\
        & LMD ($\lambda_{\text{b}}=15$) & 5.18  & \textbf{2.31}  & 1.44  & \textbf{0.84 } & \textbf{0.91 } & \textbf{1.11 } \\
        & LMD ($\lambda_{\text{b}}=0$) & 3.66  & 2.48 & 2.88  & 3.55  & 4.05  & 4.32  \\
    \midrule
    \multirow{3}[2]{*}{Black-box} & Vocoder & \textbf{33.33 } & 38.71  & 42.00  & 30.84  & 29.25  & 30.00  \\
        & LMD ($\lambda_{\text{b}}=15$) & 44.44  & 35.48  & 32.00  & 28.97  & 29.25  & 28.33  \\
        & LMD ($\lambda_{\text{b}}=0$) & 40.74  & \textbf{32.26 } & \textbf{22.00 } & \textbf{23.36 } & \textbf{23.13 } & \textbf{20.00 } \\
    \bottomrule
    \end{tabular}}
    \label{tab:realAE-exps}
\end{table}

\subsubsection{Exclusion of Failed Adversarial Examples}
\label{sssec:realAE-ablation}
In Tables \ref{tab:detectors_perf_whitebox} and \ref{tab:detectors_perf_blackbox}, the adversarial examples that are failed to attack the ASV systems are taken into the account when reporting the performance of the detectors. However, they show in fact slight difference from the genuine examples from the perspective of not only the ASV systems but also human listeners, so as to the detectors. In this section, we study how the detectors perform when we exclude the failed adversarial examples.

\tablename{\ref{tab:realAE-exps}} analyzes the performance of the detectors against the adversarial examples that can successfully attack the ASV system. The successful adversarial examples in the white-box scenario are able to move greatly away from the decision threshold, which results in an easy discrimination between adversarial and genuine examples. In contrast, the successful adversarial examples in the black-box only slightly cross the decision threshold, and thus only achieve a detection EER of 20\% at best. However, compared to \tablename{\ref{tab:detectors_perf_blackbox}}, the performance of the detectors in the black-box scenario improve substantially after excluding those failed adversarial examples.

\section{Conclusions} \label{sec:conclusion}
In this paper, we have proposed a detection-based passive defense approach called LMD to detect adversarial example for ASV systems. It is attacker-independent and possesses high interpretability. First, it masks out the regions of complex spectrograms with little speaker information to introduce a large impact on adversarial examples, and small impact on genuine examples, respectively. Then, it identifies the adversarial examples by calculating the ASV score variations before and after the masking operation. Experimental results show that our proposed LMD achieves comparable performance with the SOTA baselines. Specifically, it achieves detection EERs of no more than 5.9\% and 10.1\% on the ECAPA\_TDNN ASV and Fast-ResNet34 ASV, respectively. LMD achieves a DSR of nearly 90\% in the stringent setting of a given FAR of 1\% when encountering the BIM attacker. In addition, we evaluated the detector performance against a given SNR budget. Experimental results on the ECAPA\_TDNN ASV show that LMD outperforms the baseline approaches at a SNR budget of higher than 37dB. In an additional experiment, we find that the LMD-AIBM detector has the effect of purifying adversarial noise, which further alleviates the threat brought by the adversarial attacks.

\bibliographystyle{IEEEtran}
\bibliography{Reference}

\end{document}